\documentclass[fleqn,usenatbib]{mnras}

\usepackage{newtxtext,newtxmath}

\usepackage[T1]{fontenc}
\usepackage{ae,aecompl}

\usepackage{graphicx}	
\usepackage{amsmath}	
\usepackage{hyperref}
\usepackage{natbib}
\usepackage[caption=false]{subfig}
\usepackage{ragged2e}
\usepackage{placeins}
\usepackage{bigints} 
\usepackage{physics}
\usepackage{xcolor}
\usepackage{cleveref}
\usepackage[export]{adjustbox}
\usepackage{longtable}
\usepackage{xtab}
\usepackage{bm}

\usepackage{scalerel,tikz}
\usetikzlibrary{svg.path}
\definecolor{orcidlogocol}{HTML}{A6CE39}
\tikzset{orcidlogo/.pic={
 \fill[orcidlogocol] svg{M256,128c0,70.7-57.3,128-128,128C57.3,256,0,198.7,0,128C0,57.3,57.3,0,128,0C198.7,0,256,57.3,256,128z};
 \fill[white] svg{M86.3,186.2H70.9V79.1h15.4v48.4V186.2z}
 svg{M108.9,79.1h41.6c39.6,0,57,28.3,57,53.6c0,27.5-21.5,53.6-56.8,53.6h-41.8V79.1z M124.3,172.4h24.5c34.9,0,42.9-26.5,42.9-39.7c0-21.5-13.7-39.7-43.7-39.7h-23.7V172.4z}
 svg{M88.7,56.8c0,5.5-4.5,10.1-10.1,10.1c-5.6,0-10.1-4.6-10.1-10.1c0-5.6,4.5-10.1,10.1-10.1C84.2,46.7,88.7,51.3,88.7,56.8z};
}}
\newcommand\orcidicon[1]{\href{https://orcid.org/#1}{\mbox{\scalerel*{
\begin{tikzpicture}[yscale=-1,transform shape]
\pic{orcidlogo};
\end{tikzpicture}
}{|}}}}

\defcitealias{Mohapatra2022characterising}{M22b}
\defcitealias{Mohapatra2022VSF}{M22a}

\title[Sol vs Comp driving]{Multiphase turbulence in galactic halos: effect of the driving}

\author[Mohapatra, Federrath \& Sharma]{
Rajsekhar Mohapatra$^{\orcidicon{0000-0002-1600-7552}\,1}$\thanks{E-mail: rajsekhar.mohapatra@anu.edu.au (RM)},
Christoph Federrath$^{\orcidicon{0000-0002-0706-2306}\,1,2}$\thanks{E-mail: christoph.federrath@anu.edu.au (CF)} and 
Prateek Sharma$^{\orcidicon{0000-0003-2635-4643}\,3}$\thanks{E-mail: prateek@iisc.ac.in (PS)}
\\
$^{1}$Research School of Astronomy and Astrophysics, Australian National University, Canberra, ACT~2611, Australia\\
$^{2}$Australian Research Council Centre of Excellence in All Sky Astrophysics (ASTRO3D), Canberra, ACT~2611, Australia\\
$^{3}$Department of Physics, Indian Institute of Science, Bangalore, KA 560012, India
}

\date{Accepted XXX. Received YYY; in original form ZZZ}

\pubyear{2022}

\begin{document}
\label{firstpage}
\pagerange{\pageref{firstpage}--\pageref{lastpage}}
\maketitle

\begin{abstract}
Supernova explosions, active galactic nuclei jets, galaxy--galaxy interactions and cluster mergers can drive turbulence in the circumgalactic medium (CGM) and in the intracluster medium (ICM). 
However, the exact nature of turbulence forced by these sources and its impact on the different statistical properties of the CGM/ICM and their global thermodynamics is still unclear. 
To investigate the effects of different types of forcing, we conduct high resolution ($1008^3$ resolution elements) idealised hydrodynamic simulations with purely solenoidal (divergence-free) forcing, purely compressive (curl-free) forcing, and natural mixture forcing (equal fractions of the two components). The simulations also include radiative cooling.
We study the impact of the three different forcing modes (sol, comp, mix) on the morphology of the gas, its temperature and density distributions, sources and sinks of enstrophy, i.e., solenoidal motions, as well as the kinematics of hot ($\sim10^7~\mathrm{K}$) X-ray emitting and cold ($\sim10^4~\mathrm{K}$) H$\alpha$ emitting gas. We find that compressive forcing leads to stronger variations in density and temperature of the gas as compared to solenoidal forcing.
The cold phase gas forms large-scale filamentary structures for compressive forcing and misty, small-scale clouds for solenoidal forcing. The cold phase gas has stronger large-scale velocities for compressive forcing.  The natural mixture forcing shows kinematics and gas distributions intermediate between the two extremes, the cold-phase gas occurs as both large-scale filaments and small-scale misty clouds.
\end{abstract}

\begin{keywords}
methods: numerical -- hydrodynamics -- turbulence -- galaxies: clusters: intracluster  medium
\end{keywords}



\section{Introduction}\label{sec:introduction}
The intracluster medium (ICM), the intragroup medium (IGrM) and the circumgalactic medium (CGM) refer to the gaseous atmosphere pervading the halos of galaxy clusters, groups and individual galaxies, respectively. These media comprise of gas in multiple temperature phases, ranging from $10$--$10^7~\mathrm{K}$ \citep[for example, see][]{Bonamente2001ApJ,Fabian2003MNRAS,Salome2006A&A} and play a critical role in the evolution of their member galaxies through the cosmic baryon cycle. Feedback from supernovae and active galactic nuclei (AGNs) jets, interaction between member galaxies and with external galaxies, infall of galaxies into the group/cluster, all drive turbulence in the halo gas. This driven turbulence plays a number of key roles in the thermodynamics of these media, to name a few--(i) it seeds density fluctuations \citep{Mohapatra2021MNRAS} in the gas, which can become thermally unstable \citep{sharma2010thermal}, since denser gas cools faster (cooling rate $\propto\rho^2$, where $\rho$ is density); (ii) it converts kinetic energy from feedback processes and galaxy--galaxy interactions into thermal energy through viscous dissipation and heats the gas; and (iii) it mixes hot ($10^6$--$10^8$~$\mathrm{K}$) and cold ($10$--$10^4$~$\mathrm{K}$) components of the gas together to form intermediate-temperature gas with a short cooling time, which generates more cold gas \citep{Armillotta2016,Gronke2018,Kanjilal2021MNRAS}. Thus, understanding the properties of turbulence, such as the driving scale, the nature of the driving, in particular the fraction of solenoidal and compressive components, and the amplitude of turbulent velocities induced by the driving, is important to understand the physics of these systems. 

Many recent theoretical and numerical studies of turbulence in the CGM/ICM \cite[for example,][]{Mohapatra2019,Shi2019,Grete2020ApJ,Mohapatra2022VSF} are limited to solenoidal (divergence-free) turbulence, and mostly ignored the effects of compressive modes. The assumption of solenoidal driving is influenced by studies of incompressible/subsonic turbulence in the earth's atmosphere and oceans, because air/water are nearly incompressible, and the Boussinesq assumption ($\div{\bm{v}}=0$, where $\bm{v}$ is velocity) holds. Although the gas motions in the ICM have been observed to be subsonic \citep{hitomi2016}, the ICM/CGM is quite compressible, i.e., Mach numbers are of the order of $0.1$--$0.5$. Moreover, we observe strong converging and diverging motions in these media - such as expanding X-ray cavities around AGN-inflated bubbles \citep{Wise2007ApJ,Biava2021A&A,Tiwari2022MNRAS}, compressive waves during galaxy infall and passages \citep{Churazov2003,Churazov2021A&A}, gas sloshing and cold fronts \citep{Botteon2021A&A,Gendron-Marsolais2021ApJ,Ichinohe2021MNRAS, Ubertosi2021MNRAS,Ueda2021ApJ,Brienza2022A&A}, which indicate density contrasts and compressive motions in the ICM. Similarly, galactic outflows consisting of radially expanding motions driven by diverging flows such as stellar winds, supernova bubbles, super-bubbles and AGNs can also drive compressive motions in the disk-halo interface and the CGM \citep{Mulcahy2017A&A,Das2019ApJ,Tejos2021MNRAS}. The evolution of these compressive motions and their role in the thermodynamics of the halo gas could be very different from the divergence-free forcing studied in current literature.

Many recent studies of turbulence in the interstellar medium (ISM) have investigated the effects of compressive driving and its effects on star formation \citep{Federrath2010A&A,konstandin2012}. They find that compressive forcing leads to stronger density fluctuations (up to $3$ times larger in their standard deviation) which significantly increases the star formation efficiency, since a larger fraction of the gas becomes gravitationally unstable \citep{Federrath2012}. Compressively driven turbulence also shows a steeper velocity power spectrum \citep{Federrath2013}, which has a slope closer to Burgers turbulence with $P_v(k)\propto k^{-2}$ \citep{Burgers1948171}, rather than Kolmogorov turbulence with $P_v(k)\propto k^{-5/3}$ \citep{kolmogorov1941dissipation}, where $k$ is the wavenumber. Although these studies model supersonic turbulence in star-forming molecular clouds (which are assumed to be isothermal) and are not directly applicable to the subsonic hot plasma in the CGM/IGrM/ICM, these effects of compressive forcing could affect both the physics of these media and our interpretations of their observations. Similar to increasing the fraction of gas that becomes Jeans-unstable in molecular clouds, compressive forcing could lead to larger fractions of the gas becoming thermally unstable in the case of halo gas, since the cooling time $\propto\rho^{-1}$. These larger density fluctuations in the hot-phase gas could also affect the relation between X-ray surface brightness fluctuations and turbulent velocities, which is used to indirectly measure turbulent gas velocities in several nearby galaxy clusters  \citep{zhuravleva2014turbulent,zhuravleva2014relation,zhuravleva2018}. The steep slopes of the velocity structure functions of H$\alpha$ filaments in nearby galaxy clusters studied in \cite{Li2020ApJ} could also be affected by the nature of turbulence forcing. Hence it is important to study the effects of different types of turbulence forcing in the context of the CGM, IGrM and ICM.

Recent numerical and theoretical studies such as \cite{Shi2018MNRAS,Zhang2019MNRASa,Zhang2019MNRASb,Zhang2020MNRAS,Shi2020MNRAS} have investigated the evolution of shocks and turbulence driven by galaxy infall and cluster mergers, and their effects on the ICM. However, they have mainly focused on more global, large-scale effects of mergers, such as changes to the ICM entropy profile, contribution of turbulent pressure to non-thermal pressure in the ICM, properties of bow shocks, accretion shocks, merger shocks, etc. The current literature lacks an in-depth investigation of the effects of the turbulence driving on the multi-phase gas physics of galaxy halos.
Studies such as \citet{Ryu2008Sci} and \citet{Iapichino2012MNRAS}, using simulations and analytical modelling, respectively, have explored the role of large-scale shocks and discontinuities in driving turbulence in the ICM and the magnification of magnetic fields through the small-scale dynamo mechanism \citep[also see][]{Federrath2016JPlPh}. \cite{Porter2015ApJ} studied the effect of the turbulence driving in an isothermal ICM setup and studied the evolution of shocks, vorticity and magnetic fields. More recently, \citet{Vazza2017MNRAS}, \citet{Wittor2017MNRAS}, \citet{Wittor2020MNRAS}, and \citet{VallesPerez2021MNRAS} have studied the evolution of enstrophy, the squared amplitude of the vorticity, and its sources and sinks using cluster-scale numerical simulations.

In this study, we conduct a set of high-resolution ($1008^3$ resolution elements) local idealised simulations in a box of size $10-40~\mathrm{kpc}$. We scan the parameter space corresponding to the turbulence driving parameter $\zeta$, which quantifies the fraction of solenoidal driving. We include radiative cooling in our model. We study the effects of the driving parameter on the morphology of the gas, the density and temperature distributions and the amplitude of density fluctuations. We also determine the different sources of vorticity and the fraction of solenoidal and compressive motions at different temperatures. In order to compare with observations, we construct mock observations of the X-ray emission from hot-phase gas ($T\gtrsim10^7~\mathrm{K}$) and H$\alpha$ emission from filaments ($T\sim10^4~\mathrm{K}$). We also compare the kinematics of the two temperature phases and their dependence on the driving parameter.

This paper is organised as follows. In \cref{sec:Methods}, we introduce our model and setup. In \cref{sec:results-discussion} we present the results from our analysis and discuss their implications for the CGM/ICM. We present the shortcomings of our model and their possible effects on our results, as well as the future prospects of our work in \cref{sec:caveats-future}, and summarise and conclude in \cref{sec:Conclusion}.

\section{Methods}\label{sec:Methods}
\subsection{Model equations}\label{subsec:ModEq}
We model the CGM/ICM as a fluid using the compressible hydrodynamic (HD) equations and the ideal gas equation of state. We evolve the following equations:
\begin{subequations}
	\begin{align}
	\label{eq:continuity}
	&\frac{\partial\rho}{\partial t}+\nabla\cdot (\rho \bm{v})=0,\\
	\label{eq:momentum}
	&\frac{\partial(\rho\bm{v})}{\partial t}+\nabla\cdot (\rho\bm{v}\otimes\bm{v})+ \nabla P=\rho\bm{a},\\
	\label{eq:energy}
	&\frac{\partial E}{\partial t}+\nabla\cdot ((E+P)\bm{v})=\rho\bm{a}\cdot\bm{v}+Q-\mathcal{L},\\
	\label{eq:tot_energy}
	&E=\frac{\rho\bm{v}\cdot\bm{v}}{2} + \frac{P}{\gamma-1},
	\end{align}
\end{subequations}
where $\rho$ is the gas mass density, $\bm{v}$ is the velocity, $P=\rho k_B T/(\mu m_p)$ is the thermal pressure, $\bm{a}$ is the turbulent acceleration field that we apply, $E$ is the total energy density, $\mu$ is the mean molecular mass, $m_p$ is the proton mass, $k_B$ is the Boltzmann constant, $T$ is the temperature, $Q(\bm{x},t)$ and $\mathcal{L}(\rho,T)$ are the thermal heating and cooling rate densities respectively, and $\gamma=5/3$ is the adiabatic index. The cooling rate density $\mathcal{L}$ is given by
\begin{equation}\label{eq:cooling_function}
\mathcal{L} = n_en_i\Lambda(T),
\end{equation}
where $n_e$ and $n_i$ 
are electron and ion number densities, respectively and $\Lambda(T)$ is the temperature-dependent cooling function of \cite{Sutherland1993} corresponding to $Z_{\odot}/3$ solar metallicity. 

\subsection{Numerical methods}\label{subsec:numerical_methods}
We use a modified version of the FLASH code \citep{Fryxell2000,Dubey2008}, version~4, for our simulations. We evolve our model equations using the HLL5R Riemann solver \citep{Bouchut2007,Bouchut2010,Waagan2011}. For time integration, we use the MUSCL-Hancock scheme \citep{van1984SIAM,Waagan2009JCoPh} and a second-order reconstruction method that uses primitive variables and ensures positive values of density and internal energy. We use a uniformly spaced 3D Cartesian grid with equal size and periodic boundary conditions along all three directions. For our ICM simulations, we use a box of size $40~\mathrm{kpc}$.  For our CGM simulations, we use a box of size $10~\mathrm{kpc}$. 
We use $1008^3$ resolution elements for our parameter scan, but also include lower resolution runs with $504^3$ resolution elements for checking convergence of our results. 

We do not explicitly include viscosity in our model, instead, the viscosity is due to numerical dissipation \citep[see][for a discussion]{Benzi2008PhRvL,Federrath2011PhRvL}. The numerical viscous scale in our simulations is expected to be around $0.6~\mathrm{kpc}$, roughly at $15\Delta x$ (where $\Delta x$ is the numerical cell size), which we obtain from fitting the power spectrum model in \cite{Kriel2022MNRAS} to our velocity power spectrum. For the hot ICM, we compare the numerical viscosity with the Spitzer viscosity values from \cite{Spitzer1962pfigbook} and find it to be suppressed by a factor of $\sim600$. This is in line with expectations from recent ICM observations such as \cite{Zhuravleva2019NatAs}, who constrain the ICM viscosity to be suppressed to around $10$--$1000$ times smaller than the Spitzer viscosity.

\subsubsection{Turbulent forcing}\label{subsubsec:Turb_forcing}
We drive turbulence by exciting only large-scale modes in spectral space, and letting turbulence develop self-consistently on smaller scales. We control the amplitude of the turbulent acceleration field $\bm{a}$ such that the power is a parabolic function of $k$ and peaks at $k=2$ (for simplicity, we have dropped the wavenumber unit $2\pi/L$). We only drive modes with $1\leq\abs{\bm{k}}L/2\pi\leq3$. We use the stochastic Ornstein-Uhlenbeck (OU) process to model $\bm{a}$ with a finite autocorrelation time-scale $t_{\mathrm{turb}}$, which we have fixed to be $260~\mathrm{Myr}$ (approximately $3$ times the eddy turnover time on the driving scale) across all our ICM simulations and $100~\mathrm{Myr}$ for CGM simulations \citep{eswaran1988examination,schmidt2006numerical,Federrath2010A&A}. In steady state, the gas reaches an rms velocity of $\sim250~\mathrm{km/s}$ for our ICM-like runs \citep[consistent with observations by the][]{hitomi2016} and $\sim80~\mathrm{km/s}$ for our CGM-like runs \citep[consistent with][]{Werk2016ApJ,Faerman2017ApJ}.

To control the contribution of solenoidal and compressive driving, we introduce a parameter $\zeta$ which denotes the fraction of contribution from solenoidal driving.
The Fourier transform of the acceleration $\bm{a}$ is given by:
\begin{subequations}
\begin{align}
    \label{eq:zeta_forcing}
    &\bm{a}(\bm{k})=(1-\zeta) \Tilde{\bm{a}}_{\mathrm{comp}}(\bm{k})+\zeta \Tilde{\bm{a}}_{\mathrm{sol}}(\bm{k})\text{, where}\\
    &\Tilde{\bm{a}}_{\mathrm{comp}}(\bm{k}) = \left(\bm{k}\cdot\Tilde{\bm{a}}(\bm{k})\right)\bm{k}/k^2 \label{eq:Forc_comp}\text{, }\\
    &\Tilde{\bm{a}}_{\mathrm{sol}}(\bm{k}) = \Tilde{\bm{a}}(\bm{k})-\Tilde{\bm{a}}_{\mathrm{comp}}(\bm{k}) \label{eq:Forc_sol}\text{, }
\end{align}
\end{subequations}
where $\zeta$ is the turbulence driving parameter, $\Tilde{\bm{a}}_{\mathrm{comp}}(\bm{k})$ and $\Tilde{\bm{a}}_{\mathrm{sol}}(\bm{k})$ are the components of $\Tilde{\bm{a}}(\bm{k})$ parallel and perpendicular to the wave vector $\bm{k}$, respectively. The quantity $\Tilde{\bm{a}}(\bm{k})$ is the non-decomposed forcing term which is evolved in Fourier space using the stochastic OU process. For further details of the forcing method, we refer the reader to section~2.1 of \cite{Federrath2010A&A}. For our analysis, we have chosen three values of the driving parameter---$\zeta=0.0$ which corresponds to purely compressive (or curl-free) forcing, $\zeta=0.5$ which corresponds to natural mixture or equal contributions from the solenoidal and compressive components \footnote{Note that the ratio of compressive power to total power for $\zeta=0.5$ is $1/3$ (and not $1/2$), since in 3D we have one longitudinal (compressive) mode and two transverse (solenoidal) modes.}, and $\zeta=1.0$, which corresponds to purely solenoidal or divergence-free forcing. 

\subsubsection{Thermal heating rate and global energy balance}\label{subsubsec:therm_heating}
In all our simulations, we maintain global energy balance. This prevents the gas from undergoing runaway cooling, which is also motivated by the rarity of cooling flows in cluster observations. In order to achieve this, we inject thermal energy into the domain at a rate $Q$, such that the net thermal+kinetic energy of the system as a whole does not decrease. We also ensure the positivity of $Q$ by setting it to zero whenever the instantaneous turbulence energy injection rate is larger than the radiative cooling rate (which is rare). Mathematically, we impose the following condition at every time step:
\begin{subequations}
\begin{equation}
    Q=\max\left(0,\frac{\rho(\bm{x},t)\left(\int\mathcal{L}\mathrm{d}V-\int \rho\bm{a}\cdot\bm{v}\mathrm{d}V\right)}{\int\rho\mathrm{d}V}\right).\label{eq:thermal_heating_rate_Q}
\end{equation}
This heating rate density is proportional to the gas density of each cell, which is motivated by the several gas-density dependent heating processes, such as heating by photons and cosmic rays. We introduce the fraction $f_{\mathrm{turb}}$ which denotes the ratio of the net turbulence energy injection rate to the net cooling rate\footnote{The ratio $f_{\mathrm{turb}}$ varies as a function of time, unlike our previous studies (\citealt{Mohapatra2019,Mohapatra2022characterising}), where we fixed $f_{\mathrm{turb}}$ to a constant value and scaled $\bm{a}$ according to \cref{eq:f_turb}. Here we maintain a fixed rms velocity and not a fixed $f_{\mathrm{turb}}$.}, given as: 
\begin{equation}
f_{\mathrm{turb}}=\frac{\int\rho\bm{a}\cdot\bm{v}\mathrm{d}V}{\int\mathcal{L}\mathrm{d}V}.\label{eq:f_turb}
\end{equation}
\end{subequations}

\subsubsection{Modifications to the cooling module}\label{subsubsec:temp_ceiling_and_cool_cutoff}
\paragraph{Temperature ceiling.}\label{par:temp_ceiling} 
Since we use a periodic box for our setup and do not let the energy escape, the thermal heat that we inject to impose global energy balance and the dissipation of turbulence kinetic energy can heat the hot phase ($T\gtrsim10^7~\mathrm{K}$ for the ICM and $T\gtrsim10^6~\mathrm{K}$ for the CGM) to unrealistic high temperatures ($T\gtrsim10^8~\mathrm{K}$). As we do not consider gravity, this hot, low-density gas does not buoyantly rise up, expand and cool as it would in a realistic stratified atmosphere. Very hot grid cells lead to an increase in the maximum speed of sound ($c_s$) and a decrease in the code time step ($dt_{\mathrm{cs}}=\Delta x
/c_s$, where $c_s=\sqrt{\gamma k_BT/\mu m_p}$, and $\Delta x$ is the size of a cell). Thus, in order to prevent the gas from reaching unrealistic hot temperatures and to control the code time step, we introduce a temperature ceiling to our cooling module, which limits any gas with $T>T_{\mathrm{ceiling}}$ to $T_{\mathrm{ceiling}}$. For all our simulations, we set $T_{\mathrm{ceiling}}=3\times10^8~\mathrm{K}$ for our ICM-like runs and $10^8~\mathrm{K}$ for our CGM-like runs. We have ensured that adding a temperature ceiling does not significantly affect the temperature distribution of the gas, but only cuts off its high temperature tail at $T_{\mathrm{ceiling}}$.

\paragraph{Pressure and temperature cutoffs on cooling.}\label{par:cool_cutoffs}
We set the cooling function to zero below a temperature $T_{\mathrm{cutoff}}$ and a pressure $P_{\mathrm{cutoff}}$. The cooling function from \cite{Sutherland1993} that we use is truncated at $10^4\,\mathrm{K}$, which we set to be our $T_{\mathrm{cutoff}}$. In order to limit numerical instability issues associated with low-temperature gas dropping to extremely low pressures, we set $P_{\mathrm{cutoff}}=P_0/1000$, where $P_0$ is the initial pressure of the gas. This cutoff only limits the gas pressure of a small portion of the cold gas from decreasing below $P_{\mathrm{cutoff}}$. We have verified that it does not significantly affect any other properties of the gas, such as density, temperature and velocity distributions.

\paragraph{Cooling subcycling.}\label{par:cooling_subcycling}
The cooling time of the gas is given by:
\begin{equation}\label{eq:tcool}
    t_{\mathrm{cool}}=\frac{E_{\mathrm{int}}}{\mathcal{L}}=\frac{P}{(\gamma-1)n_e n_i\Lambda(T)}.
\end{equation}
The cooling time can be quite short for gas at intermediate temperatures ($10^5~\mathrm{K}\lesssim T\lesssim10^6~\mathrm{K}$) where $\Lambda(T)$ peaks. This makes our multiphase turbulence runs numerically expensive. In order to achieve a reasonable run-time, we evolve the internal energy with operator splitting in the short $t_{\mathrm{cool}}$ grid cells using smaller time steps. We reduce the global timestep if it exceeds the shortest cooling time by a fixed factor $\mathrm{sub}_{\mathrm{factor}}$. We have explained this module in detail in section~2.2.5 of  \cite{Mohapatra2022characterising} (hereafter \citetalias{Mohapatra2022characterising}). In this setup, we set the hydrodynamical time step to
\begin{equation}\label{eq:cool_subcycling}
    \mathrm{dt}_{\mathrm{code}}=\mathrm{min}(0.5\times \mathrm{sub}_{\mathrm{factor}} \times t_{\mathrm{cool,min}},\mathrm{dt}_{\mathrm{CFL}}),    
\end{equation}
where $t_{\mathrm{cool,min}}$ is the minimum value of $t_{\mathrm{cool}}$ in the entire domain and $\mathrm{dt}_{\mathrm{CFL}}$ is the time step set by the CFL (Courant-Friedrichs-Lewy) criterion. We use $\mathrm{sub}_{\mathrm{factor}}=25$ for this study. We have tested the effect of different values of $\mathrm{sub}_{\mathrm{factor}}$ on the temperature distribution of the gas and discuss this further in Appendix~C.

\paragraph{Density ceiling on cooling.}\label{par:dens_ceiling_cooling}
The cooling time $t_{\mathrm{cool}}\propto\rho^{-1}$. For compressive driving runs, the cold phase can reach high densities due to the combined effects of cooling and convergent driving. In order to control the code time step, we set the cooling function to zero for gas with density $\rho>\rho_{\mathrm{ceiling}}$. In this study, we use $\rho_{\mathrm{ceiling}}=500\times\rho_0$, where  $\rho_0$ is the initial density. We have checked the effect of introducing a density ceiling on the density distribution of the gas and find that it only affects a small part of the high-density tail (c.f. fig. \ref{fig:dens-PDF}).

Thus, the complete cooling function is given by
\begin{equation}
    \mathcal{L}=n_en_i\Lambda(T)\mathcal{H}(T-T_{\mathrm{cutoff}})\mathcal{H}(P-P_{\mathrm{cutoff}})\mathcal{H}(\rho_{\mathrm{ceiling}}-\rho), \label{eq:cooling_modified}
\end{equation}
where $\mathcal{H}$ is the Heaviside function.

\subsection{Initial conditions}\label{subsec:init_conditions}
For our ICM-like runs, we use the same initial conditions as in our setup in \citetalias{Mohapatra2022characterising}. We initialise the gas with a temperature $T_0=4\times10^6~\mathrm{K}$, $n_e=0.086~\mathrm{cm}^{-3}$ and with an initial sound speed of $300~\mathrm{km/s}$. These conditions mimic the cool central regions of groups and clusters. For our CGM setup, we initialise the gas with $T_0=10^5~\mathrm{K}$, but with the same initial density as the ICM-like runs.
We drive turbulence in the box using the same forcing module (with the same seed for the OU process) for all three runs. We only vary $\zeta$ across the three runs and control the amplitude of the driving, such that in steady state the gas has similar root-mean-square (rms) velocities across our different $\zeta$ runs. 
This driven turbulence creates density fluctuations in the gas, which can cool in a runaway fashion. 
The gas separates into hot and cold phases, with the cold phase at $T_{\mathrm{cutoff}}=10^4~\mathrm{K}$ and the hot phase at $10^7$--$10^8~\mathrm{K}$, which mimics typical ICM conditions.

\subsection{List of simulations}\label{subsec:list_of_models}
\begin{table*}
	\centering
	\def\arraystretch{1.2}
	\caption{Simulation parameters for different runs}
	\label{tab:sim_params}
	\resizebox{\textwidth}{!}{
		\begin{tabular}{lcccccccc} 
			\hline
			Label & Driving & Resolution  & $\mathrm{sub}_{\mathrm{factor}}$ & $\mathcal{M}_{\mathrm{hot}}$ & $v\ (\mathrm{km/s})$  &   $v_{\mathrm{comp}}\ (\mathrm{km/s})$ & $\sigma_{s,\mathrm{hot}}^2$  & $\sigma_{\ln(\bar{P}),\mathrm{hot}}^2$ \\
			(1) & (2) & (3) & (4) & (5) & (6) & (7) & (8) & (9)\\
			\hline
			$\zeta0.0$ & Compressive & $1008^3$ & $25$ & $0.20\pm0.01$ & $250\pm30$ & $100\pm30$ & $0.36\pm0.04$ & $0.012\pm0.005$ \\
			$\zeta0.5$ & Natural     & $1008^3$ & $25$ & $0.23\pm0.01$ & $240\pm30$ & $90\pm20$ & $0.23\pm0.01$ & $0.006\pm0.001$ \\
			$\zeta1.0$ & Solenoidal  & $1008^3$ & $25$ & $0.24\pm0.01$ & $240\pm30$ & $90\pm20$ & $0.19\pm0.01$ & $0.006\pm0.002$ \\
		    \hline
			$\zeta0.0$LR & Compressive & $504^3$ & $25$ & $0.18\pm0.01$ & $240\pm20$ & $80\pm20$ & $0.34\pm0.02$ & $0.008^{+0.003}_{-0.002}$\\
			$\zeta0.5$LR & Natural     & $504^3$ & $25$ & $0.23\pm0.01$ & $240\pm30$ & $70\pm10$ & $0.22\pm0.01$ & $0.0045\pm0.0008$\\
			$\zeta0.5$LR$10$ & Natural     & $504^3$ & $10$ & $0.27\pm0.02$ & $250\pm30$ & $80\pm10$ & $0.19\pm0.01$ & $0.010^{+0.004}_{-0.003}$ \\
			$\zeta0.5$LR$1$ & Natural     & $504^3$ & $1$ & $0.32\pm0.02$ & $250\pm30$ & $90\pm20$ & $0.15\pm0.02$ & $0.025^{+0.013}_{-0.008}$ \\
			$\zeta1.0$LR & Solenoidal  & $504^3$ & $25$ & $0.23\pm0.01$ & $230\pm30$ & $60\pm10$ & $0.18\pm0.01$ & $0.0033^{+0.008}_{-0.006}$\\
			\hline
			$\zeta0.0$LR-CGM & Compressive & $504^3$ & $25$ &  $0.14\pm0.02$ & $80\pm10$ & $32\pm10$ & $0.57\pm0.09$ & $0.003^{+0.004}_{-0.001}$\\
			$\zeta0.5$LR-CGM & Natural     & $504^3$ & $25$ &  $0.28\pm0.04$ & $80\pm10$ & $33\pm7$ & $0.26^{+0.14}_{-0.09}$ & $0.02^{+0.03}_{-0.01}$\\
			$\zeta1.0$LR-CGM & Solenoidal  & $504^3$ & $25$ &  $0.24\pm0.02$ & $70\pm10$ & $27\pm4$ & $0.34\pm0.05$ & $0.008\pm0.002$\\
		    \hline
			$\zeta0.0$HR-nocool & Compressive & $1536^3$ & N/A & $0.20\pm0.01$ & $270\pm30$ & $140\pm50$ & $0.059\pm0.005$ & $0.066\pm0.005$ \\
			$\zeta0.5$HR-nocool & Natural     & $1536^3$ & N/A & $0.25\pm0.01$ & $270\pm20$ & $16\pm4$ & $0.0011\pm0.0001$ & $0.0026\pm0.0004$ \\
			$\zeta1.0$HR-nocool & Solenoidal  & $1536^3$ & N/A & $0.24\pm0.01$ & $240\pm30$ & $5\pm1$ & $0.0005\pm0.0001$ & $0.0010\pm0.0002$ \\
		    \hline
			
	\end{tabular}}
	\justifying \\ \begin{footnotesize} Notes: Column 1 shows the simulation label. The number following $\zeta$ denotes the fraction of power in solenoidal modes in our turbulent driving. We denote the type of driving in column 2. In column 3, we show the resolution of the simulations. We use $1008^3$ cells for our main parameter study, but we have also performed simulations at lower resolution ($504^3$, indicated by `LR' in the label) for checking convergence. The fourth column shows the subcycle factor for a simulation, defined in \cref{eq:cool_subcycling}. The default value of $\mathrm{sub}_{\mathrm{factor}}$ is $25$, unless indicated in the simulation label (following LR). In column 5, we show the volume-weighted rms Mach number of the hot phase ($T>10^7~\mathrm{K}$for ICM-like runs and $T>10^6~\mathrm{k}$ for CGM-like runs). In columns 6 and 7, we show the volume-weighted standard deviations of velocity $v$ and of the longitudinal component of velocity $v_{\mathrm{comp}}$, respectively ($v^2=v_{\mathrm{comp}}^2+v_{\mathrm{sol}}^2$). Finally, in columns 8 and 9, we show $\sigma^2_{s,\mathrm{hot}}$ and $\sigma^2_{\ln{\bar{P}},\mathrm{hot}}$, the squares of the standard deviations of the logarithms of density and pressure of the hot phase, respectively.\end{footnotesize} 
	
\end{table*}

We scan the parameter space of the driving parameter $\zeta$ using three high-resolution simulations ($1008^3$ resolution elements) with $\zeta$ values $0.0$ (compressive forcing), $0.5$ (natural mixture) and $1.0$ (solenoidal forcing). These simulations use ICM-like initial conditions and form our fiducial set, which we study in most of our analysis. To extend the applicability of our results to smaller galactic halos, we have conducted three simulations with $\zeta$ values $0.0$, $0.5$ and $1.0$ and CGM-like initial conditions, and present them in \cref{subsec:CGM case}. We further conduct three non-radiative simulations with $\zeta$ values $0.0$, $0.5$ and $1.0$, and present their results in Appendix~A, which helps to distinguish between the effects of driving and the effects of radiative cooling and thermal instability. In order to check convergence with numerical resolution, we have conducted three more ICM-like simulations at a resolution of $504^3$ grid cells. We discuss the dependence of our results on resolution in Appendix~B. We also test the dependence on the sub-cycling factor $\mathrm{sub}_{\mathrm{factor}}$ in Appendix~C, and we have conducted two more simulations with $\mathrm{sub}_{\mathrm{factor}}=10\text{ and }1$. These simulations, along with some of their key setup parameters and statistical properties are listed in \cref{tab:sim_params}. 

\section{Results and discussion}\label{sec:results-discussion}
In this section, we present the key results of our work and discuss their implications to the physics of the halo gas and our interpretation of their observations. We start with the results from our high-resolution ICM-like runs and study the dependence of their gas properties on the driving parameter. We then discuss the effect of the solenoidal driving fraction $\zeta$ on interpreting X-ray and H$\alpha$ observations of the ICM. Finally, we present the effect of $\zeta$ on the gas properties of our CGM-like runs.

\subsection{2D projection and slice}\label{subsec:2D_proj}
\begin{figure*}
		\centering
	\includegraphics[width=2.0\columnwidth]{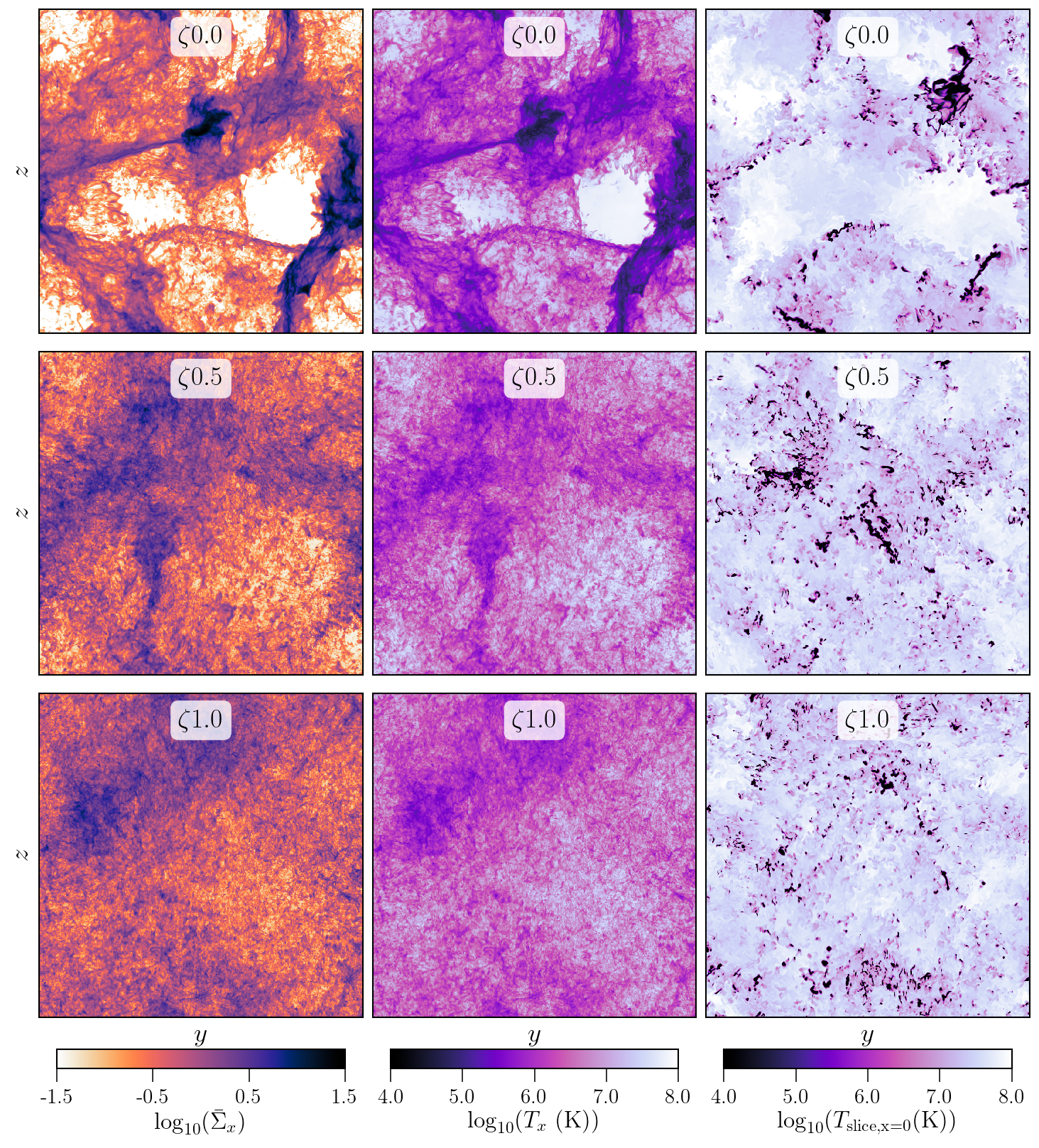}	
	\caption[density-temp-projection plots]{First column: snapshots of normalised density projected along the $x$ direction at $t=1.042$~$\mathrm{Gyr}$ for three representative simulations: $\zeta0.0$, $\zeta0.5$ and $\zeta1.0$. Second column: mass-weighted projections of temperature at this snapshot along the $x$ direction for these runs. Third column: A slice of temperature in the $yz$ plane, at $x=0$. All colorbars are in log scale, and are shown at the bottom of each column. These plots show strong large-scale variations in the gas density and temperature for the $\zeta0.0$ run and more small-scale, weaker variations for the $\zeta0.5$ and $\zeta1.0$ runs.}
	\label{fig:dens-temp-proj-2d}
\end{figure*}

In \cref{fig:dens-temp-proj-2d} we show projections of the logarithm of volume-weighted gas density and mass-weighted temperature integrated along the $x$-direction ($\Sigma_x$ and $T_x$, respectively) and a slice of temperature along the $yz$-plane at $x=0$ at $t=1.042~\mathrm{Gyr}$. Across all three runs, we find a clear correlation between the density and the temperature of the gas---dense regions correspond to cooler gas at $\sim10^4~\mathrm{K}$ and rarefied regions correspond to hot gas with $T\sim10^7$--$10^8~\mathrm{K}$. 

With increasing $\zeta$, we observe a change in the morphology of the gas. The contrast between high- and low-density regions (or corresponding cold- and hot-phase regions), denoted by $\chi=\rho_{\mathrm{cold}}/\rho_{\mathrm{hot}}$ (where $\rho_{\mathrm{cold}}$ and $\rho_{\mathrm{hot}}$ denote the characteristic gas densities of the cold and hot phases, respectively), is much stronger for the compressive forcing run ($\zeta0.0$) and this decreases for the natural ($\zeta0.5$) and solenoidal forcing ($\zeta1.0$) runs. This is a result of strong converging and diverging motions associated with compressive driving.
For the $\zeta0.0$ run, the cold phase has a filamentary structure and it surrounds the low-density hot-phase regions, seen in both the temperature projection and slice. The cold clouds are large in size, almost equivalent to the driving scale ($L/2=20~\mathrm{kpc}$). This is also the scale on which we expect the density fluctuations to be the strongest. However, these structures are transient, lasting around a few $10\mathrm{s}$--$100~\mathrm{Myr}$ (see movie of projected temperature in supplementary data). Although the cold-phase gas does shatter to small scales, we observe a lack of misty small-scale structures. The turbulent mixing time scale $t_{\mathrm{mix}}\propto\ell/u_\ell$, decreases with decreasing $\ell$ (for both compressive and solenoidal turbulence). The lack of small-scale clouds for compressive forcing indicates that large-scale modes dominate the evolution of the medium. Turbulent mixing on small scales is relatively weak, which is expected due to the larger $\chi$ (density contrast) for this run \citep[also see section~4.4 of][]{Gronke2022MNRAS}. The small-scale clouds are swept up to form high-density filaments due to our continuous compressive forcing. As we increase $\zeta$, $\chi$ decreases due to the lack of strong compressive modes. The number of high-density large-scale cold clouds decreases. This also leads to enhanced turbulent mixing on small scales and the production of a large number of misty small-scale cold clouds. Thus, large-scale cold structures can shatter and be mixed with the ambient hot phase and we require continuous compressive forcing to generate and maintain them. 


\subsection{Evolution of cold gas mass, rms Mach number and longitudinal velocities}\label{subsec:time-evolution}

\begin{figure}
		\centering
	\includegraphics[width=\columnwidth]{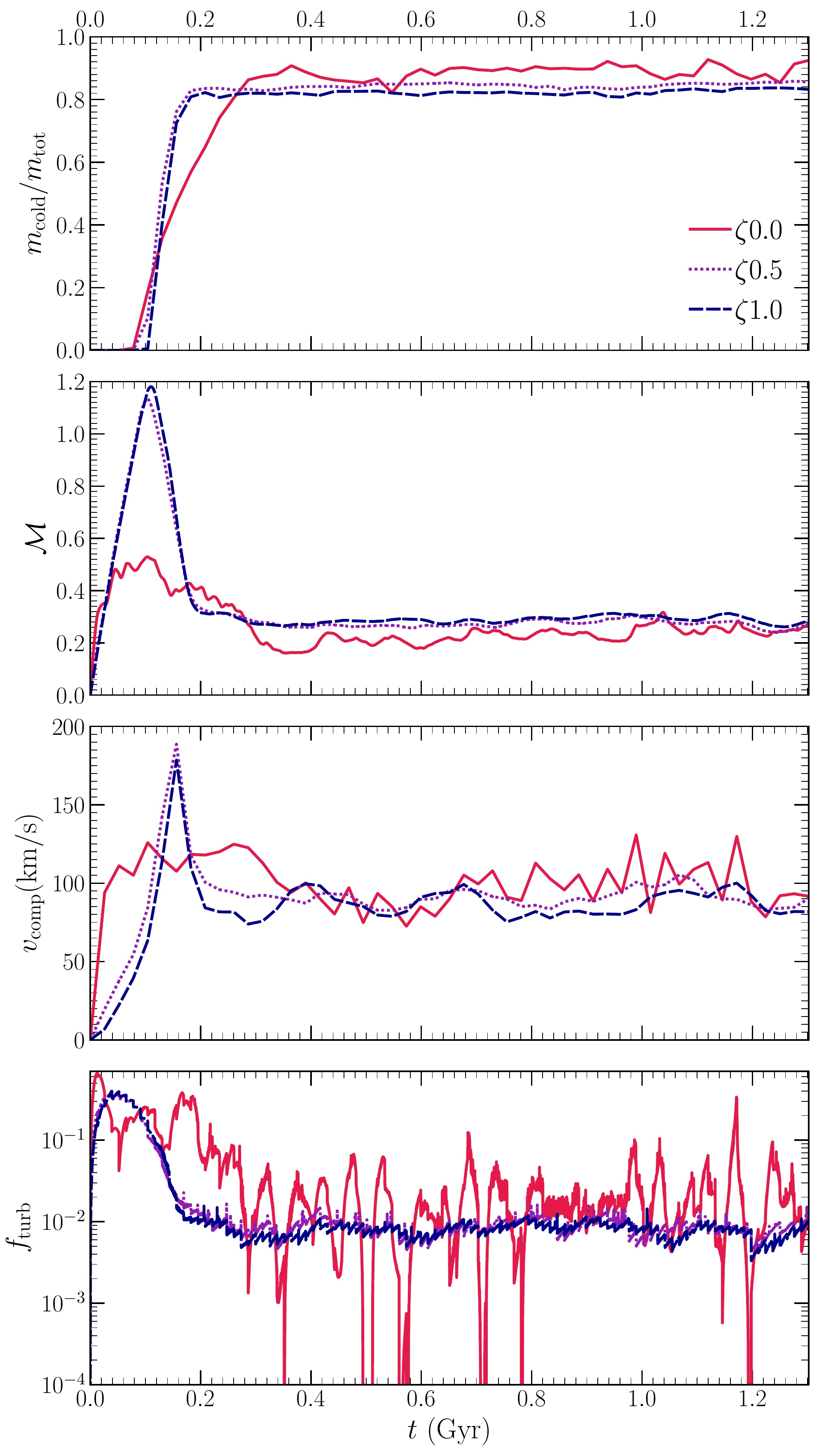}	
	\caption[time-evolution plots]{Time evolution of mass fraction of gas in the cold phase (first panel $m_\mathrm{cold}/m_{\mathrm{tot}}$), volume-weighted rms Mach number (second panel $\mathcal{M}$), the volume-weighted standard deviation of the longitudinal component of velocity (third panel, $v_{\mathrm{comp}}$) and the turbulence power fraction (fourth panel, $f_{\mathrm{turb}}$, defined in \cref{eq:f_turb}). In general, the $\zeta0.5$ run shows properties closer to the $\zeta1.0$ run. The $\zeta0.0$ run shows the strongest temporal variation in all the above quantities.
	}
	\label{fig:time-evolution}
\end{figure}

In this subsection, we discuss the evolution of some statistical properties of our multiphase setup. In the first panel of \cref{fig:time-evolution}, we show the evolution of net mass fraction of cold gas $m_{\mathrm{cold}}/m_{\mathrm{tot}}$, where $m_{\mathrm{cold}}$ refers to the total mass of gas with temperature $T<2\times10^4~\mathrm{K}$ (the results are insensitive to the exact choice of the cut-off temperature as long as the PDF peak at $10^4$ K is captured). Initially, the gas is at $T_0=4\times10^6~\mathrm{K}$, so $m_{\mathrm{cold}}/m_{\mathrm{tot}}=0$. As we drive turbulence, we seed density fluctuations in the gas, where the denser regions cool faster than the lower-density regions. All three runs start forming multiphase gas around $100~\mathrm{Myr}$ and then the fraction $m_{\mathrm{cold}}/m_{\mathrm{tot}}$ reaches a steady state value of $\sim0.8$--$0.9$ at around $\sim200$--$300~\mathrm{Myr}$. We expect compressive forcing to generate stronger density fluctuations compared to solenoidal forcing, where the densest regions have a shorter cooling time. Due to these fast cooling seed over-densities, the $\zeta0.0$ run forms multiphase gas the fastest among the three. The $\zeta0.5$ run forms multiphase gas at almost the same time as the $\zeta0.0$ run, and its later evolution closely follows that of the $\zeta1.0$ run. The fraction $m_{\mathrm{cold}}/m_{\mathrm{tot}}$ shows a sharp increase once the initial seed over-densities start cooling. It reaches the steady state value faster for the $\zeta0.5$ and $\zeta1.0$ runs compared to the $\zeta0.0$ run. The later evolution of $m_{\mathrm{cold}}/m_{\mathrm{tot}}$ is controlled by small-scale solenoidal modes, which form cold gas through turbulent mixing of the hot and cold phases. The solenoidal modes saturate (i.e., reach a balance between generation and dissipation) later for the lower $\zeta$ runs, which we discuss later in \cref{subsec:evol_enstrophy_sources}. 
In steady state, the fraction $m_{\mathrm{cold}}/m_{\mathrm{tot}}$ decreases with increasing $\zeta$. The larger value of this fraction is associated with the high-density cold gas structures in \cref{fig:dens-temp-proj-2d}, which become less abundant with increasing $\zeta$. 

In the second and third panels of \cref{fig:time-evolution}, we show the evolution of the volume-weighted rms Mach number $\mathcal{M}$ and the volume-weighted standard deviation of the longitudinal component of velocity $v_{\mathrm{comp}}$. Both $\mathcal{M}$ and $v_{\mathrm{comp}}$ for the $\zeta0.0$ run increase rapidly initially and reach a steady state value. We do not find any strong correlations between the evolution of $\mathcal{M}$ or $v_{\mathrm{comp}}$ with $m_{\mathrm{cold}}/m_{\mathrm{tot}}$, especially around the time when the cold gas forms. In contrast, for the $\zeta0.5$ and $\zeta1.0$ runs, $\mathcal{M}$ and $v_{\mathrm{comp}}$ increase initially, then show a steep increase around $t\sim100~\mathrm{Myr}$, marked by a change in the slope. 
This is followed by a decrease in the values of $\mathcal{M}$ and $v_{\mathrm{comp}}$ which reach their steady state values by $300~\mathrm{Myr}$. The sharp initial increase corresponds to strong compressive flows associated with the condensation of the dense cold-phase gas and the decrease later corresponds to the dissipation of these compressive modes.  
In steady state, the value of $v_{\mathrm{comp}}$ decreases slightly with increasing $\zeta$ (see column~7 of \cref{tab:sim_params} for its temporally averaged value in steady state). This is expected, since compressive velocities in the $\zeta0.0$ run are driven directly in addition to the converging flows associated with gas condensation.

In the fourth panel of \cref{fig:time-evolution}, we show the evolution of $f_{\mathrm{turb}}$, which denotes the ratio of the kinetic energy injection rate due to external turbulence forcing and the net radiative cooling rate, defined in \cref{eq:f_turb}. For all three runs, $f_{\mathrm{turb}}$ is positive and in steady state it has a value of around $1\%$. This is in line with our results from \citetalias{Mohapatra2022characterising}, where we found that weaker turbulence driving runs best reproduced ICM-like gas temperature and Mach number distributions. 

The average value of $f_{\mathrm{turb}}$ is the largest for the $\zeta0.0$ run. This run also shows the strongest variations in $f_{\mathrm{turb}}$. In general, the injected energy dissipates faster for compressive modes\footnote{The turbulence energy injection rate for the $\zeta0.0$ run is $8$ times larger than the rate for the $\zeta1.0$ run, while they produce similar $v$. In steady state, assuming that the injection rate is balanced by the dissipation rate, we infer that the compressive modes dissipate faster than the solenoidal modes. This inference is corroborated by the quicker saturation of $v_{\mathrm{comp}}$ in the third row of \cref{fig:time-evolution}.} (at around the sound crossing time on the driving scale $\sim L/2c_s$) compared to solenoidal modes (which dissipate over a few $t_{\mathrm{eddy}}\sim L/2u_L$ on the driving scale). Although the value of $\mathcal{M}_{\mathrm{hot}}$ is smaller for the $\zeta0.0$ run, $f_{\mathrm{turb}}$ is larger because we inject kinetic energy at a faster rate for this run compared to the $\zeta0.5$ and $\zeta1.0$ runs for similar turbulent velocity amplitudes. 

\subsection{Density and temperature probability distribution functions}\label{subsec:dens_temp_PDFs}
\begin{figure}
		\centering
	\includegraphics[width=\columnwidth]{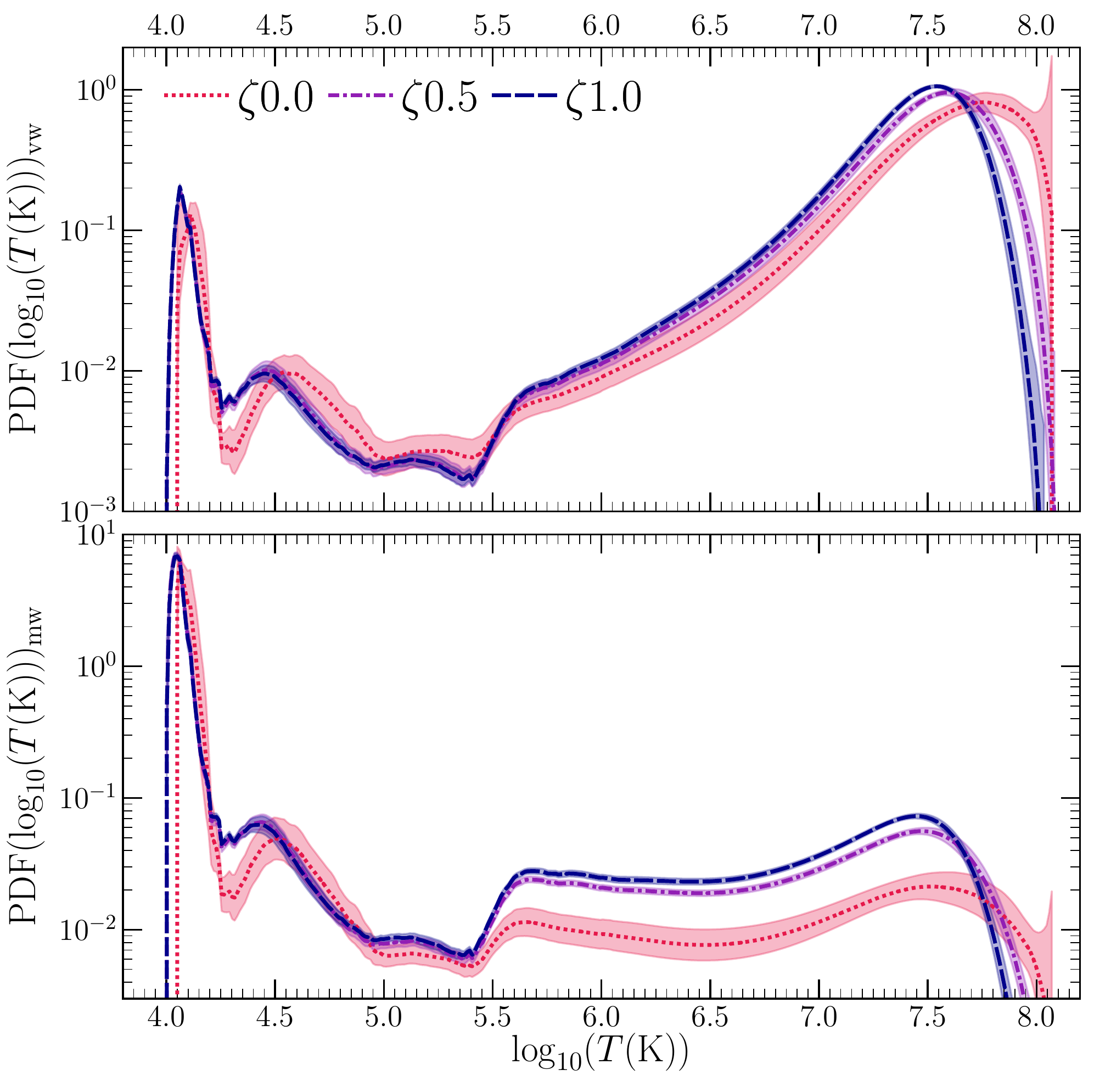}	
	\caption[temperature PDF]{Upper panel: Volume-weighted PDF of temperature for our three representative runs. Lower panel: Mass-weighted PDFs of temperature for the same runs. With increasing $\zeta$, there is more gas (by mass) at $T\gtrsim10^{5.5}\,\mathrm{K}$ and the temperature of the hot-phase peak decreases.}
	\label{fig:temp-PDF}
\end{figure}

Here we present the volume-weighted probability distribution functions (PDF) of gas temperature and density for our fiducial runs and discuss their implications. We average the PDFs temporally from $t=0.65~\mathrm{Gyr}$ to $1.3~\mathrm{Gyr}$, which is when turbulence is fully developed and is roughly in a steady state.

\subsubsection{Temperature distribution}\label{subsubsec:temp_distribution}
\begin{figure}
		\centering
	\includegraphics[width=\columnwidth]{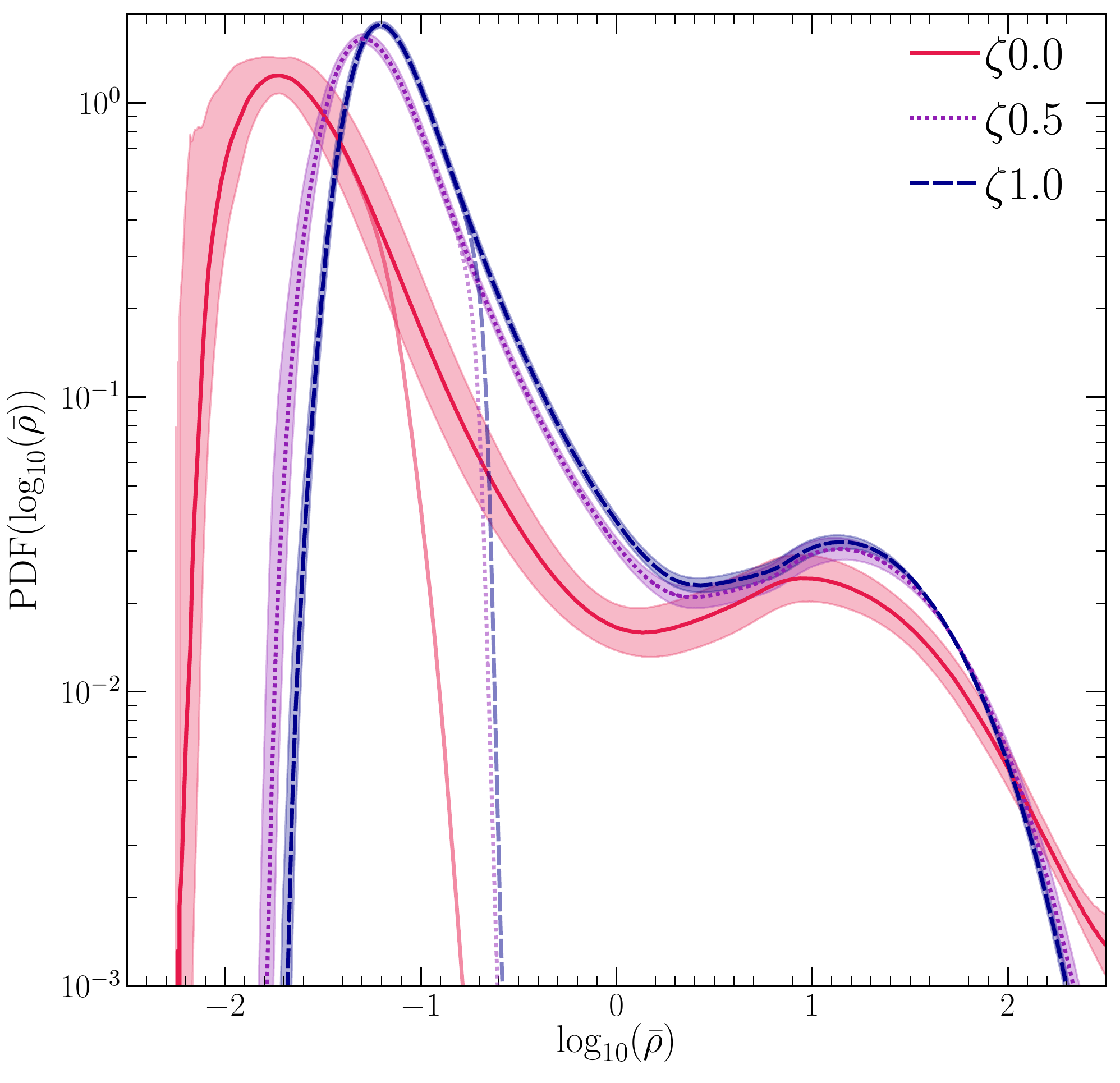}	
	\caption[density PDF]{Volume-weighted PDF of density for our three representative runs with different fractions of solenoidal and compressive forcing. We average the PDFs in steady state from $t=0.65~\mathrm{Gyr}$ to $1.3~\mathrm{Gyr}$. The shaded regions around the PDFs show the $1-\sigma$ temporal variation in the PDF. The lighter coloured lines show the PDFs of gas in the hot phase ($T>10^7$~$\mathrm{K}$). The density bimodality decreases with increasing solenoidal fraction $\zeta$.}
	\label{fig:dens-PDF}
\end{figure}

We show the volume-weighted temperature PDF in the upper panel of \cref{fig:temp-PDF}. For all three runs, the gas has two distinct peaks, one close to $T_{\mathrm{cutoff}}~(10^4~\mathrm{K})$ and another between $10^7$--$10^8~\mathrm{K}$. There is a clear lack of gas at intermediate temperatures between the two peaks. The small bumps at these intermediate temperatures correspond to the shape of the cooling curve. With increasing $\zeta$ (the solenoidal driving parameter), the temperature of the hot phase decreases slightly. In the lower panel of \cref{fig:temp-PDF}, we show the mass-weighted temperature PDF. The amplitude of the cold-phase peak is much larger than that of the hot-phase peak. The mass of gas with $10^{5.5}\lesssim T/\mathrm{K}\lesssim 10^8$ increases With increasing $\zeta$. This implies that for the run with compressive driving, the hot gas has lower density compared to the runs with solenoidal or natural driving. This feature is also seen in \cref{fig:dens-PDF}, which we discuss below.

\subsubsection{Density distribution}\label{subsubsec:dens_distribution}
In \cref{fig:dens-PDF} we show the PDF of the logarithm of density ($\log_{10}(\bar{\rho})$, where $\bar{\rho}=\rho/\rho_0$). The shaded region shows the $1\sigma$ variation in time. The lighter coloured lines show the PDF when we only consider the hot-phase gas ($T>10^7~\mathrm{K}$). In all three cases, the density PDF has two peaks, where the low-density peak corresponds to the hot phase ($\rho_{\mathrm{hot}}$) and the high-density peak corresponding to the cold phase ($\rho_{\mathrm{cold}}$). The value of $\chi$ decreases with increasing $\zeta$. This is explained by strong converging and diverging motions associated with compressive forcing, which lead to large variations in gas density. The $\zeta0.0$ run also shows a high-density tail, which extends till $\bar{\rho}\approx10^{2.5}$, due to compressive motions further compressing the cold dense regions and increasing its density. These high-density power-law tails are also seen in the density distribution of expanding HII region shocks  \cite[][see their fig.~16]{Tremblin2012A&A} as well as compressively forced turbulence simulations of the warm and cold neutral media in \cite{Seifried2011A&A}.  The amount of gas at intermediate densities, in between $\rho_{\mathrm{cold}}$ and $\rho_{\mathrm{hot}}$ increases with increasing $\zeta$. This can we explained by weaker mixing between the hot and cold phases due to the larger $\chi$ values associated with the compressive forcing run. 

\subsection{Phase diagram of density and temperature fluctuations}\label{subsec:dens_temp_joint_PDF}

\begin{figure*}
		\centering
	\includegraphics[width=2.0\columnwidth]{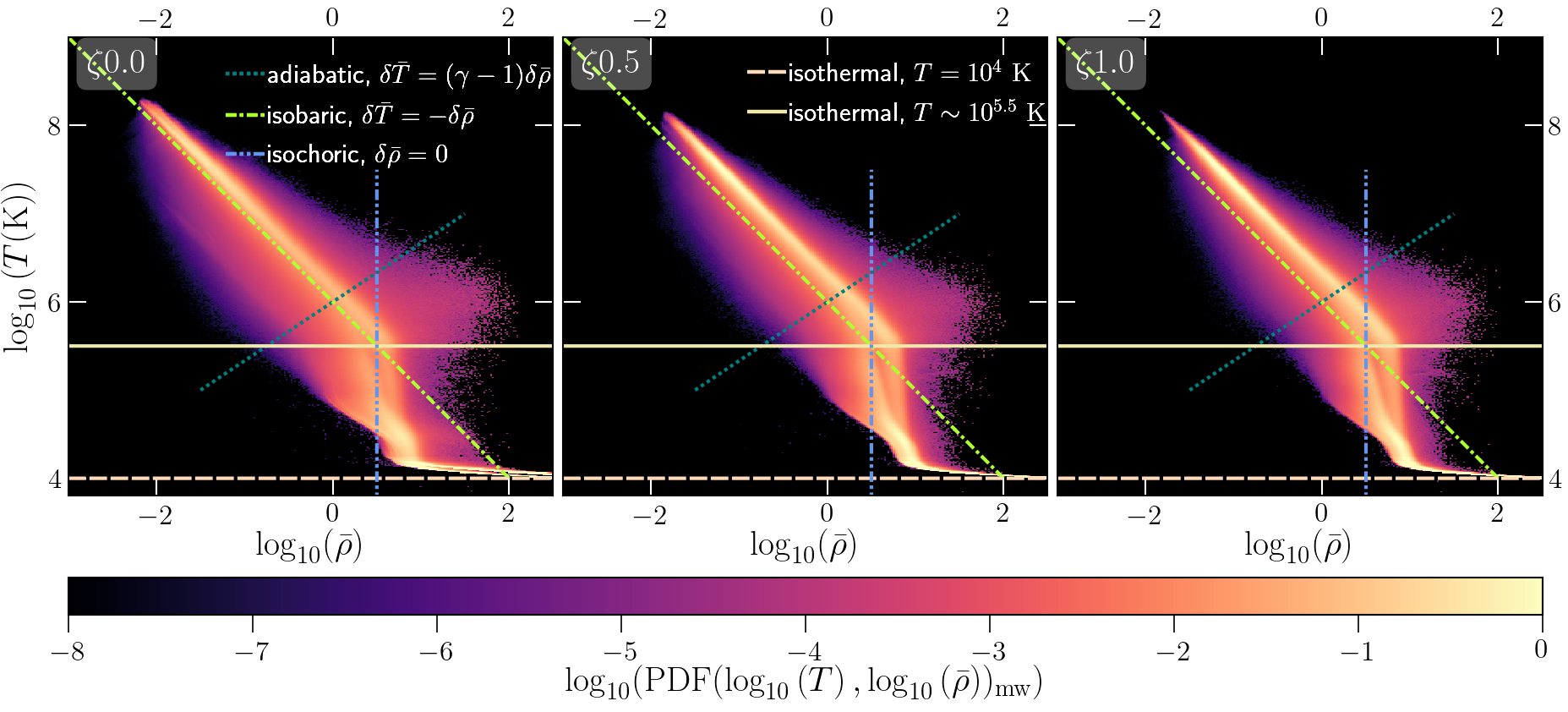}	
	\caption[Temperature vs density 2D PDF]{The two-dimensional mass-weighted PDF of the logarithm of temperature versus logarithm of density for our three fiducial runs averaged over $2t_{\mathrm{eddy}}$, from $t=0.78~\mathrm{Gyr}$ to $t=0.94~\mathrm{Gyr}$. The straight lines show the different fluctuation modes--adiabatic, isobaric, isothermal,  and isochoric. The gas with $T\gtrsim10^{5.5}~\mathrm{K}$ is mostly isobaric, and isochoric just below $\sim10^{5.5}~\mathrm{K}$--which corresponds to the peak of cooling function that we use, where $t_{\mathrm{cool}}\ll t_{\mathrm{cs}}$ across the cooling blob. The cold gas accumulates just above the isothermal track at $T_{\mathrm{cutoff}}=10^4$~$\mathrm{K}$. We do not observe any strong trends with varying $\zeta$.
	\label{fig:temp-dens-mw-2D-PDF}
	}
\end{figure*}

We show the joint mass distribution of $\log_{10}{T}$ and $\log_{10}{\bar{\rho}}$ in \cref{fig:temp-dens-mw-2D-PDF}, averaged from $t=0.78~\mathrm{Gyr}$ to $t=0.94~\mathrm{Gyr}$. The straight lines show the different fluctuation modes--adiabatic, isobaric, isochoric and isothermal. The gas fluctuations are isobaric for $T\gtrsim10^{5.5}~\mathrm{K}$ and below this temperature, the gas cooling to lower temperatures is mainly isochoric for our three fiducial runs. This temperature corresponds to the peak of the \cite{Sutherland1993} cooling curve that we use, where $t_{\mathrm{cool}}\ll t_{\mathrm{cs}}$ across the cooling blob. The trends that we observe in \cref{fig:temp-dens-mw-2D-PDF} agree with the weak turbulent heating run (f0.001) in \citetalias{Mohapatra2022characterising} (see their fig.~8), since $f_{\mathrm{turb}}\sim10^{-2}$ for all three runs (fourth panel of \cref{fig:time-evolution}). We do not observe any strong effects of the nature of driving on these phase diagrams.  Our results are in agreement with the X-ray observations by \cite{Zhuravleva2016MNRAS,zhuravleva2018}, who find perturbations in the hot gas to be mostly isobaric.

\subsection{Density and pressure fluctuations in the hot phase gas} \label{subsec:dens_pres_fluc_hot}

\begin{figure*}
		\centering
	\includegraphics[width=1.5\columnwidth]{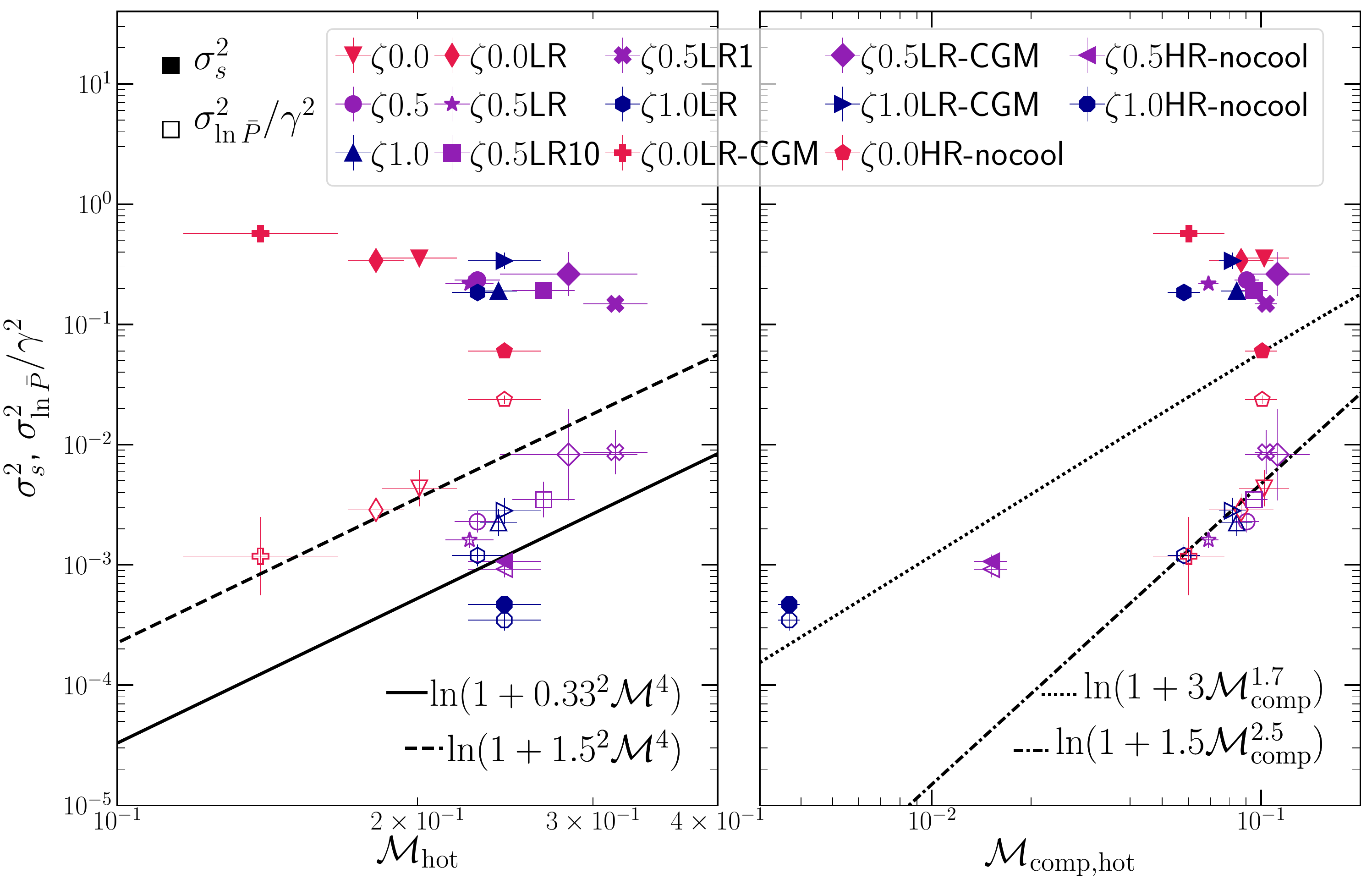}	
	\caption[hot gas density and pressure fluctuations]{\underline{Left panel:} The variance of $s$ (filled data points) and $\ln(\bar{P})$ (unfilled), the logarithm of normalised density and pressure, respectively of the hot-phase gas (defined as $T>10^6$~$\mathrm{K}$ for CGM-like runs, $T>10^7$~$\mathrm{K}$ for ICM-like runs) plotted as a function of the rms Mach number of the hot phase $\mathcal{M}_{\mathrm{hot}}$ (left panel) . The solid and dashed line shows the predicted scaling relation for subsonic turbulence with solenoidal forcing (coefficient $b=0.33$) without radiative cooling (see eq.~13f in \citealt{Mohapatra2021MNRAS} with the Froude number $\mathrm{Fr}\rightarrow\infty$). We choose $b=1.5$ for compressive forcing, which seems to agree well with the scaling of the pressure fluctuations. In general, the pressure fluctuations agree better with the scaling relation as compared to density fluctuations, which are much larger in the presence of radiative cooling due to the associated thermal instability rather than purely driven by turbulence. \underline{Right panel:} Similar to the left panel, except we plot the fluctuations (squared) against the compressive rms Mach number of the hot phase $\mathcal{M}_{\mathrm{comp,hot}}$ along the $x$-axis. The dotted line shows the scaling relation from \cite{konstandin2012}, which agrees relatively well with the amplitude of density and pressure fluctuations in our non-radiative (nocool) runs. The hot phase pressure fluctuations in radiative runs show a much steeper scaling, varying as $\mathcal{M}_{\mathrm{comp,hot}}^{2.5}$, where as the density fluctuations are larger than the scaling relation and do not show any dependence on $\mathcal{M}_{\mathrm{comp,hot}}$. 
	\label{fig:sig-mach-hot}
	}
\end{figure*}

The hot-phase density follows a log-normal distribution for all three runs, as seen in \cref{fig:dens-PDF}. We list the values of $\sigma_{s,\mathrm{hot}}^2$ (where $s=\ln(\bar{\rho})=\log_{10}(\bar{\rho})/\ln(10)$) and $\sigma_{\ln{\bar{P}},\mathrm{hot}}^2$ (the width of the pressure PDF\footnote{We multiply $\sigma_{\ln{\bar{P}}}^2$ by $1/\gamma^2$ so that for subsonic and non-stratified turbulence without cooling, it has similar scaling with $\mathcal{M}$ as $\sigma_s^2$ \citep[see][]{Mohapatra2021MNRAS}.}) in columns~8 and~9 of \cref{tab:sim_params}, respectively. In \cref{fig:sig-mach-hot}, we show $\sigma_s^2$ and $\sigma_{\ln{\bar{P}}}^2/\gamma^2$ as a function of the hot-phase rms Mach number $\mathcal{M}_{\mathrm{hot}}$ (left panel) and the hot-phase compressive rms Mach number $\mathcal{M}_{\mathrm{comp,hot}}$ (right panel).
The amplitude of the density and pressure fluctuations in the hot phase are used to obtain indirect estimates of turbulent gas velocities using both X-ray surface brightness \citep{zhuravleva2014turbulent,zhuravleva2014relation,Eckert2017ApJ,Bonafede2018MNRAS} and thermal Sunyaev-Zeldovich effect (tSZ) observations \citealt{khatri2016}, respectively \citep[see][for a review]{Simionescu2019SSRv}.

We observe that the amplitude of the density fluctuations is clearly much larger than the scaling relation between $\sigma_s^2$ and $\mathcal{M}_{\mathrm{hot}}$ from \cite{Mohapatra2021MNRAS} (for unstratified, non-radiative sims, shown as solid and dashed lines in the left panel). The value of $\sigma_{s,\mathrm{hot}}^2$ decreases with increasing $\mathcal{M}_{\mathrm{hot}}$ which may be due to enhanced turbulent mixing of large-density fluctuations in these runs. We also compare the value of $\sigma_{s,\mathrm{hot}}^2$ to the scaling relation for subsonic isothermal turbulence in \cite{konstandin2012} in the right panel, where the relation of $\sigma_{s,\mathrm{hot}}^2$ with $\mathcal{M}_{\mathrm{comp,hot}}$ is shown using the dotted line. We find that density fluctuations for the non-radiative runs agree with this scaling relation and for the radiative runs, they are slightly larger than the prediction for the corresponding $\mathcal{M}_{\mathrm{comp,hot}}$. 

\cite{Simonte2022A&A} find the density fluctuations to be independent of the degree of stratification of the ICM in their sample of galaxy clusters, with $\sigma_s^2\sim0.05$--$0.5$. This is consistent with the \cite{konstandin2012} scaling relation for $\mathcal{M}_{\mathrm{comp,hot}}$ in the range $0.1$--$0.4$. The amplitude of the density fluctuations are also slightly larger than those predicted by \cite{Mohapatra2020,Mohapatra2021MNRAS}, who only considered solenoidal driving. Since the fluctuations generated by compressive motions are much larger, it explains the lack of dependence of the fluctuations on the degree of stratification in \cite{Simonte2022A&A}.

Thus, the density structure in radiative runs is shaped mainly by thermal instability of the medium (and not the turbulence driving), which generates large compressive velocities and contact discontinuities leading to larger fluctuations.
X-ray surface brightness fluctuations, which measure the amplitude of gas density fluctuations may overestimate the turbulent gas velocities in thermally unstable regions of the clusters and galactic halos, if one uses the $\sigma_s$--$\mathcal{M}$ scaling relation calibrated using non-radiative simulations. However, they can be useful to obtain an upper limit on the compressive component of the velocity field.
 
In comparison to the density fluctuations, the pressure fluctuations are much smaller in amplitude and appear to scale with the $\mathcal{M}_{\mathrm{hot}}$ as $\sigma_{\ln{\bar{P}}}\propto\mathcal{M}_{\mathrm{hot}}^2$ for constant $\zeta$ (left panel of \cref{fig:sig-mach-hot}), for both radiative and non-radiative (nocool) runs. But the constant of proportionality is larger for smaller value of $\zeta$ ($\sim1.5^2$ for compressive forcing, $\sim0.33^2$ for solenoidal), as expected for stronger compressive forcing. This dependence on $\zeta$ is quantified better in the plot on the right panel of \cref{fig:sig-mach-hot}, where we find that $\sigma_{\ln{\bar{P}}}^2\propto\mathcal{M}_{\mathrm{comp,hot}}^{2.5}$ for the radiative runs, where as non-radiative runs do not show this scaling. The exact scaling of $\sigma_{\ln{\bar{P}}}$ is likely to depend on both $\mathcal{M}_{\mathrm{hot}}$ and $\mathcal{M}_{\mathrm{comp,hot}}$, but this is not the focus of our current study and we leave a detailed analysis of the scaling relation for future work.  

Thus, in thermally unstable regions of the halo gas, the large X-ray brightness fluctuations could be due to contact discontinuities and compressive motions associated with thermal instability, and not be associated with turbulence. In these regions, tSZ fluctuations are more robust probes of both turbulent gas velocities and their compressive component. One could separate the effects of the two through a Helmholtz decomposition of the observed quantity.

\subsection{Compressive to solenoidal velocity ratio}\label{subsec:comp_sol_ratio}

\begin{figure*}
		\centering
	\includegraphics[width=2.0\columnwidth]{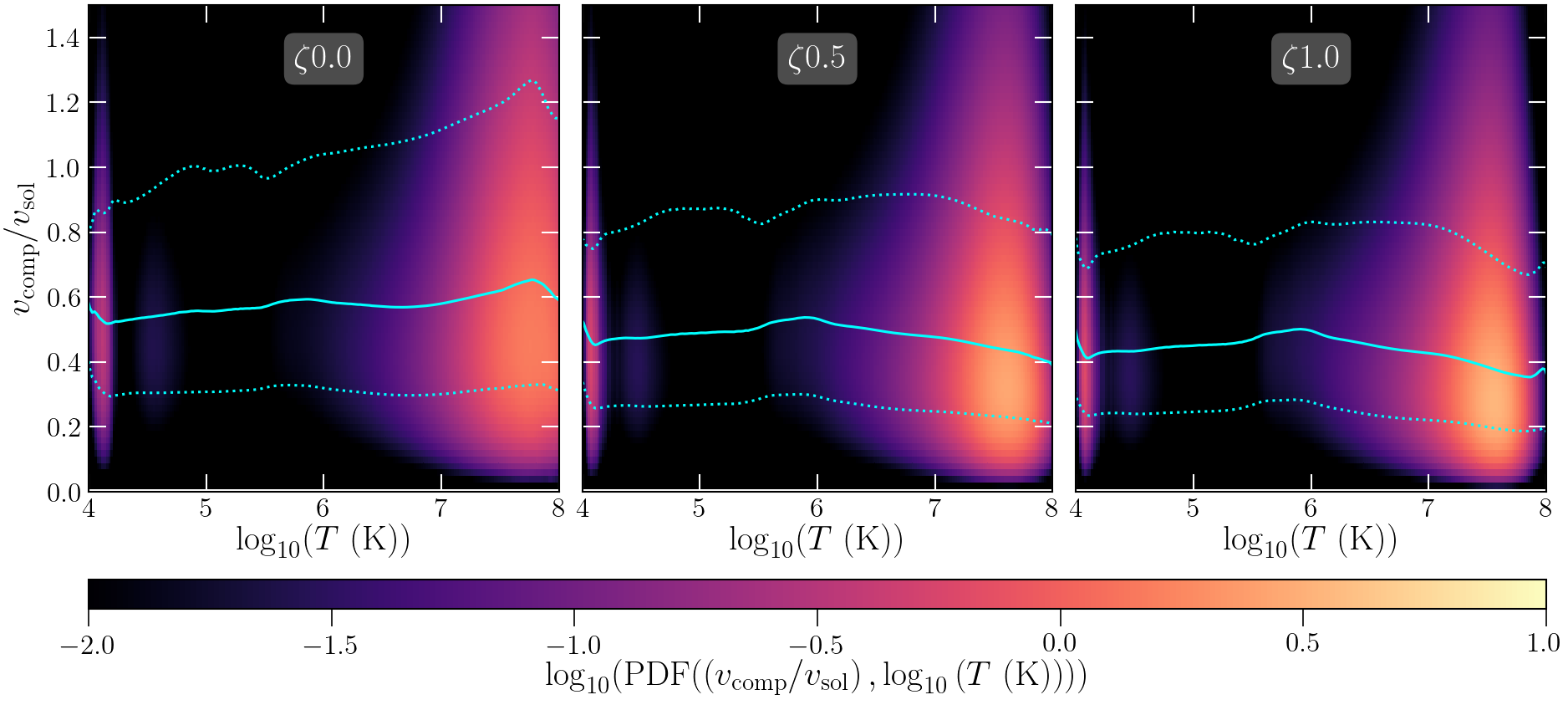}	
	\caption[comp-temp-2d-pdf]{Two-dimensional volume-weighted probability distribution functions of the ratio between compressive to solenoidal velocity magnitudes $v_{\mathrm{comp}}/v_{\mathrm{sol}}$. The solid cyan line shows the median distribution ($50\mathrm{th}$ percentile) of this ratio as a function of the temperature, and the dotted lines show the $16\mathrm{th}$ and $84\mathrm{th}$ percentiles. Although the ratio increases with decreasing $\zeta$, its median value is $\sim0.6$ for $\zeta0.0$, indicating the dominance of solenoidal modes for all the runs.}
	\label{fig:comp-temp-pdf-2d}
\end{figure*}

We calculate the compressive ($\bm{v}_{\mathrm{comp}}$) and solenoidal ($\bm{v}_{\mathrm{sol}}$) components of the velocity field by performing a Helmholtz decomposition of the velocity field ($\bm{v}$) in Fourier space, followed by an inverse Fourier transform of the decomposed $\bm{v}$-components. This methodology for the Helmholtz decomposition of the acceleration field is described in \cref{eq:Forc_comp} and \cref{eq:Forc_sol}. In \cref{fig:comp-temp-pdf-2d}, we present the ratio between the compressive and solenoidal velocity amplitudes $v_{\mathrm{comp}}/v_{\mathrm{sol}}$ and its dependence on the temperature of the gas in a 2D volume-weighted PDF. 
For all three runs, the hot phase ($T>10^7~\mathrm{K}$) has the strongest peak in the PDF since it is the volume-filling phase. It shows the broadest distribution in the value of  $v_{\mathrm{comp}}/v_{\mathrm{sol}}$. Although the velocities are predominantly solenoidal, a significant fraction of the hot gas also has a stronger compressive velocity component. At intermediate temperatures, we have very less gas and the gas motions are predominantly solenoidal with a small spread. The cold-phase gas ($10^4~\mathrm{K}<T<2\times10^4~\mathrm{K}$) shows a large distribution in the value of the ratio, including regions where the compressive component of the velocity is stronger than the solenoidal component. 

We study the variation of the ratio $v_{\mathrm{comp}}/v_{\mathrm{sol}}$ with temperature and its dependence on $\zeta$ using the cyan lines. The solid cyan line shows the median value of the ratio as a function of temperature. The dotted cyan lines show the $16^{\mathrm{th}}$ and $84^{\mathrm{th}}$ percentiles of the ratio, respectively. The median value of the ratio for all three fiducial runs is much smaller than the energy equipartition value of $1/\sqrt{2}$ (due to one compressive mode and two transverse modes for each $\bm{k}$ in 3D). Even with compressive forcing, the gas velocities are predominantly solenoidal. To further investigate this result, we look into the different sources of enstrophy $\epsilon$ (which gives us the power in solenoidal modes) in the next subsection. 

With increasing $\zeta$, the median value of the ratio $v_{\mathrm{comp}}/v_{\mathrm{sol}}$ decreases slightly across all temperatures. This decrease is much stronger for the hot phase compared to the cold phase. The ratio does not depend strongly on $\zeta$ for the cold and intermediate temperature gas--since intermediate gas is mainly formed by mixing of hot- and cold-phase gas through small-scale solenoidal modes and cold-phase gas forms subsequently through fast cooling of this gas. The dynamics on small scales is dominated by solenoidal modes and is largely unaffected by the large-scale driving. 

In a recent study, \cite{Choudhury2022arXiv} have modelled gentle AGN feedback by injecting thermal energy in the centre of their simulated cluster. 
They find that the energy fraction of solenoidal modes is an order of magnitude larger than the compressive modes, similar to our results. However, the compressive modes are comparable in terms of energy dissipation rate. Solenoidal turbulence  dissipation dominates on small scales, closer to the cluster centre, whereas large-scale compressive modes contribute significantly to the heating of the ICM away from the bubbles. \cite{Vazza2017MNRAS} model cluster formation using large-scale cosmological simulations and they find turbulence dissipation to be dominated by solenoidal motions, especially near the cluster centre. The contribution of compressive modes towards heating the ICM becomes important on larger scales and is time-dependent, for example in cluster outskirts and during merger events.

\subsection{Evolution of enstrophy and its sources}\label{subsec:evol_enstrophy_sources}

\begin{figure}
		\centering
	\includegraphics[width=\columnwidth]{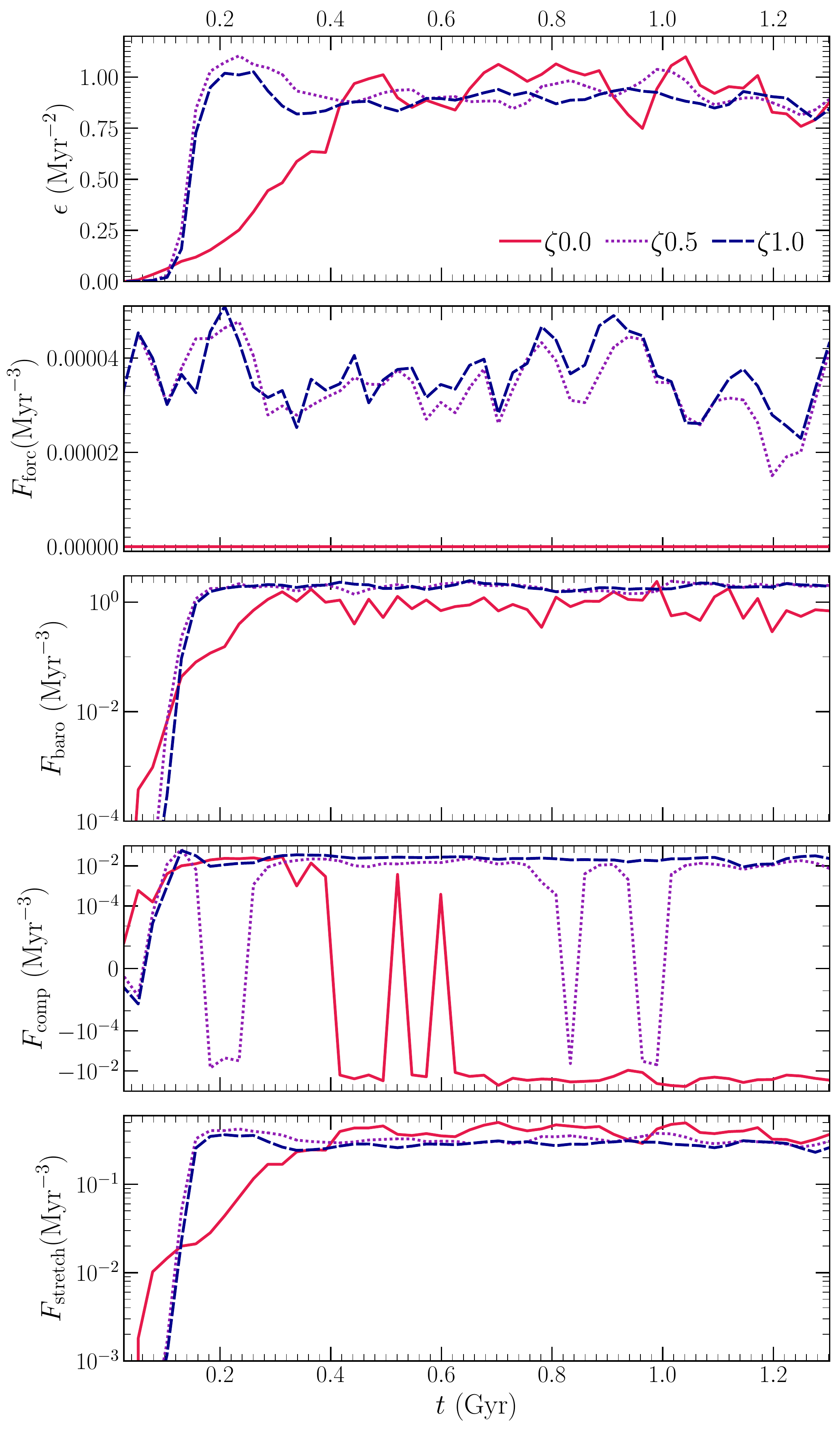}	
	\caption[enstrophy sources evolution]{The evolution of volume-averaged enstrophy ($\epsilon$) and its sources for our three fiducial runs. Note that in the fourth panel, the $y$-axis is set to be linear between $\pm10^{-4}~\mathrm{Myr}^{-3}$ and logarithmic otherwise. In steady state, all three runs have similar amplitudes of enstrophy. Among the source terms, the baroclinic and vortex stretching terms are dominant, whereas the forcing and compression terms are orders of magnitude smaller.}
	\label{fig:enstrophy-sources-evolution}
\end{figure}

In the previous subsection, we observed that the gas velocities are predominantly solenoidal in nature, even for our compressive forcing run. In this subsection, we look for the sources and sinks of enstrophy, which is often used as a proxy for solenoidal turbulence. We derive the evolution equation for the enstrophy ($\epsilon=|\nabla\times\bm{v}|^2/2$) by dividing the momentum equation (\cref{eq:momentum}) by $\rho$, taking its curl, followed by taking a dot product with $\bm{\omega}$ (where $\bm{\omega}=\nabla\times\bm{v}$ is the vorticity). The equation for the evolution of enstrophy is given by
\begin{subequations}
\begin{align}
    \label{eq:enstrophy_evolution}
     \frac{\partial\epsilon}{\partial t} = &-\bm{\nabla}\cdot(\bm{v}\epsilon)-\epsilon\bm{\nabla}\cdot\bm{v}+\bm{\omega}\cdot(\bm{\omega}\cdot\bm{\nabla})\bm{v}\;+\\
     &\quad\frac{\bm{\omega}}{\rho^2}\cdot(\bm{\nabla}\rho\times\bm{\nabla}P)+\bm{\omega}\cdot\left(\bm{\nabla}\times\bm{a}\right).\nonumber
\end{align}
We denote the source terms using the same notations as previous studies \citep[such as][]{Porter2015ApJ,Vazza2017MNRAS} for easier comparison. 
The different enstrophy source terms are then defined as
\begin{align}
    &F_{\mathrm{adv}}=-\bm{\nabla}\cdot(\bm{v}\epsilon)\text{, corresponding to net enstrophy inflow;}\label{eq:enstrophy_adv}\\
    &F_{\mathrm{comp}}=-\epsilon\bm{\nabla}\cdot\bm{v}\label{eq:enstrophy_comp} \text{, corresponding to increase in enstrophy} \\
    \nonumber &\text{with fluid contraction;}\\
    &F_{\mathrm{baro}}=\frac{\bm{\omega}}{\rho^2}\cdot(\bm{\nabla}\rho\times\bm{\nabla}P)\label{eq:enstrophy_baro}\text{, corresponds to the baroclinic term,}\\
    \nonumber  &\text{i.e., misalignments between density and pressure gradients;}\\
    &F_{\mathrm{stretch}}=\bm{\omega}\cdot(\bm{\omega}\cdot\bm{\nabla})\bm{v} \label{eq:enstrophy_stretch} \text{, corresponding to enstrophy} \\
    \nonumber &\text{generation due to stretching of vortices;}\\
    &F_{\mathrm{forc}}=\bm{\omega}\cdot\left(\bm{\nabla}\times\bm{a}\right)\text{, corresponding to enstrophy generation} \label{eq:enstrophy_forcing} \\
    \nonumber &\text{due to external turbulence forcing.}
\end{align}
\end{subequations}

Among all these terms, only the forcing term and the baroclinic term can generate vorticity when its initial value is zero. \footnote{Note that we have ignored the contribution of the dissipation term  $F_{\mathrm{diss}}=\nu\bm{\omega}\cdot\left(\nabla^2\bm{\omega}+\bm{\nabla}\times\bm{G}\right)$ to the enstrophy evolution, where $\nu$ is kinematic viscosity and $\bm{G}=(1/\rho)\bm{\nabla}\rho\cdot\bm{S}$, $\bm{S}$ denotes the trace-less strain tensor. Although we do not have an explicit viscosity term, numerical viscosity could still act as a source of enstrophy. The second term can act as a source of vorticity from its zero initial value \citep[see][paragraph after eq.~2 for a discussion]{Federrath2011PhRvL}. We cannot directly calculate the contribution of $F_{\mathrm{diss}}$ due to its numerical origin, but we can estimate its relative importance in steady state (when $\partial\epsilon/\partial t\approx0$) by taking the sum of all the other source terms.} This is particularly relevant to our $\zeta0.0$ runs with compressive forcing, where $\bm{\nabla}\times\bm{a}=0$ and the initial vorticity is zero. 


In \cref{fig:enstrophy-sources-evolution}, we show the evolution of volume-averaged $\epsilon$ and the source terms $F_{\mathrm{forc}}$, $F_{\mathrm{baro}}$, $F_{\mathrm{comp}}$ and $F_{\mathrm{stretch}}$ for our three fiducial runs. The advection term $F_{\mathrm{adv}}$ is zero due to our periodic boundary conditions.
For all runs, $\epsilon$ initially increases as a function of time, eventually reaching a steady state. The $\zeta0.0$ run takes much longer to reach the steady-state value. This delay in reaching a steady state is also seen in \cref{fig:time-evolution}, where the cold-gas mass fraction in the $\zeta0.0$ run saturates later than in the other two runs. It is caused by weak vortex stretching ($F_{\mathrm{stretch}}$) and baroclinic  terms ($F_{\mathrm{baro}}$) (see fifth and third panels of \cref{fig:enstrophy-sources-evolution}) at initial times. Since we do not drive solenoidal modes for this run, the seeds for enstrophy growth through vortex stretching are smaller. But $F_{\mathrm{stretch}}$ slowly grows to reach a similar value as the higher $\zeta$ runs. Although the value of $F_{\mathrm{baro}}$ at $t=0.06~\mathrm{Gyr}$ is larger than the other two, after multiphase gas formation (see first row of \cref{fig:time-evolution}) it grows slower due to a lower number of cold-hot phase boundaries, where baroclinicity is strong. The value of $\epsilon$ in steady state is also similar for our three fiducial runs. Even in steady state, the value of $\epsilon$ shows relatively large fluctuations for the $\zeta0.0$ run, which is a characteristic of the compressive modes (e.g., see evolution of $v_{\mathrm{comp}}$ in third panel of \cref{fig:time-evolution}).

\subsubsection{Forcing}\label{subsubsec:forcing_enstrophy} As expected, the rate of enstrophy generation due to external turbulence forcing  $F_{\mathrm{forc}}=0$ for the $\zeta0.0$ run, since forcing is curl-free ($\bm{\nabla}\times\bm{a}=\bm{0}$). Although the value of $F_{\mathrm{forc}}$ increases with increasing $\zeta$, it is of the order of $\sim\text{a few}\times10^{-5}~\mathrm{Myr}^{-3}$ and is significantly smaller than some of the other source terms.


\subsubsection{Baroclinicity}\label{subsubsec:baroclinicity_enstrophy} The baroclinic source term $F_{\mathrm{baro}}$ (third panel of \cref{fig:enstrophy-sources-evolution}) is the largest source term and is always positive. At initial times, $F_{\mathrm{baro}}$ is the largest for the $\zeta0.0$ run, which acts as the seed of enstrophy, since all other source terms are zero initially (barring the contribution due to $F_{\mathrm{diss}}$, which we consider to be negligible at initial times). At later times ($t\gtrsim0.5~\mathrm{Gyr}$), the gas separates into hot and cold phases, leading to an even stronger and positive $F_{\mathrm{baro}}\sim\text{a few}\times10^{-2}~\mathrm{Myr}^{-3}$ for all three runs. In this stage, $F_{\mathrm{baro}}$ is larger for the higher $\zeta$ runs. This is due to the larger number of small cold-phase clouds along whose boundaries we expect the baroclinicity to be strong.

\subsubsection{Compression}\label{subsubsec:compression_enstrophy} The compression term $F_{\mathrm{comp}}$ (fourth panel,  \cref{fig:enstrophy-sources-evolution}) is two orders of magnitude smaller than the two largest terms ($F_{\mathrm{baro}}$ and $F_{\mathrm{stretch}}$). For the $\zeta0.5$ and $\zeta1.0$ runs, $F_{\mathrm{comp}}$ is mostly positive, implying the compressive motions associated with cloud condensation generate enstrophy. On the other hand, $F_{\mathrm{comp}}$ is negative for the $\zeta0.0$ run due to expansive motions associated with the large-scale driving, which reduces (or leads to smaller relative growth of) the enstrophy. 

\subsubsection{Vortex stretching}\label{subsubsec:vort_stretch_enstrophy} The vortex stretching term $F_{\mathrm{stretch}}$ (fifth panel, \cref{fig:enstrophy-sources-evolution}) is always positive and is comparable in amplitude to $F_{\mathrm{baro}}$. Vorticity, seeded by $F_{\mathrm{forc}}$ and $F_{\mathrm{baro}}$ is further amplified by $F_{\mathrm{stretch}}$, similar to small-scale turbulent dynamo amplification \citep[see for e.g.][]{Federrath2016JPlPh,Seta2020MNRAS}. 

\subsubsection{Comparison with recent studies}\label{subsubsec:comparison_enstrophy}

\cite{Wittor2020MNRAS,Wittor2021MNRASErratum} study the evolution of enstrophy in their cluster simulation. 
They find that all enstrophy source terms ($F_{\mathrm{stretch}}$, $F_{\mathrm{adv}}$, $F_{\mathrm{comp}}$ and $F_{\mathrm{baro}}$) show temporal correlation with AGN activity \citep[fig.~2 in][]{Wittor2021MNRASErratum}. In contrast to our results, in their study $F_{\mathrm{baro}}$ is the weakest source term. It is comparable to the other terms only during periods of AGN activity, implying that the misaligned density and pressure gradients could be short-lived for subsonic ICM turbulence. Compressive forcing and cold gas condensation (like in this study) may be important to generate them.

\citet{Vazza2017MNRAS}, \citet{Wittor2017MNRAS}, and \citet{VallesPerez2021MNRAS} study the evolution of enstrophy and its sources in samples of clusters from cosmological simulations. They find that enstrophy is initially generated at the outermost accretion shocks by $F_{\mathrm{baro}}$ and $F_{\mathrm{comp}}$, as well as  by other shocks during cluster evolution. In agreement with our results, $F_{\mathrm{stretch}}$ is one of the strongest terms at all times. It distributes enstrophy from strong shocks into unshocked regions. They also find that turbulent gas motions are predominantly solenoidal, although the driving is mainly compressive (through inner and merger shocks), similar to our results in \cref{subsec:comp_sol_ratio}. 

\cite{Seta2022arXiv} study the effect of the driving parameter on the turbulent dynamo in the two-phase ISM. Similar to our study, they find $F_{\mathrm{baro}}$ to generate seeds of enstrophy in their compressive driving run at initial times. The cross term in $F_{\mathrm{diss}}$ also plays a role in seeding enstrophy, as we discussed in the footnote. 
The Lorentz force acts as an additional source of enstrophy in their magnetohydrodynamic setup, however it is subdominant in both their dynamo growth and saturation phase, as compared to $F_{\mathrm{stretch}}$ and $F_{\mathrm{baro}}$ which are the largest source terms in steady state.


\subsection{2D mock-emission maps}\label{subsec:2D_emission}
\begin{figure*}
		\centering
	\includegraphics[width=2.0\columnwidth]{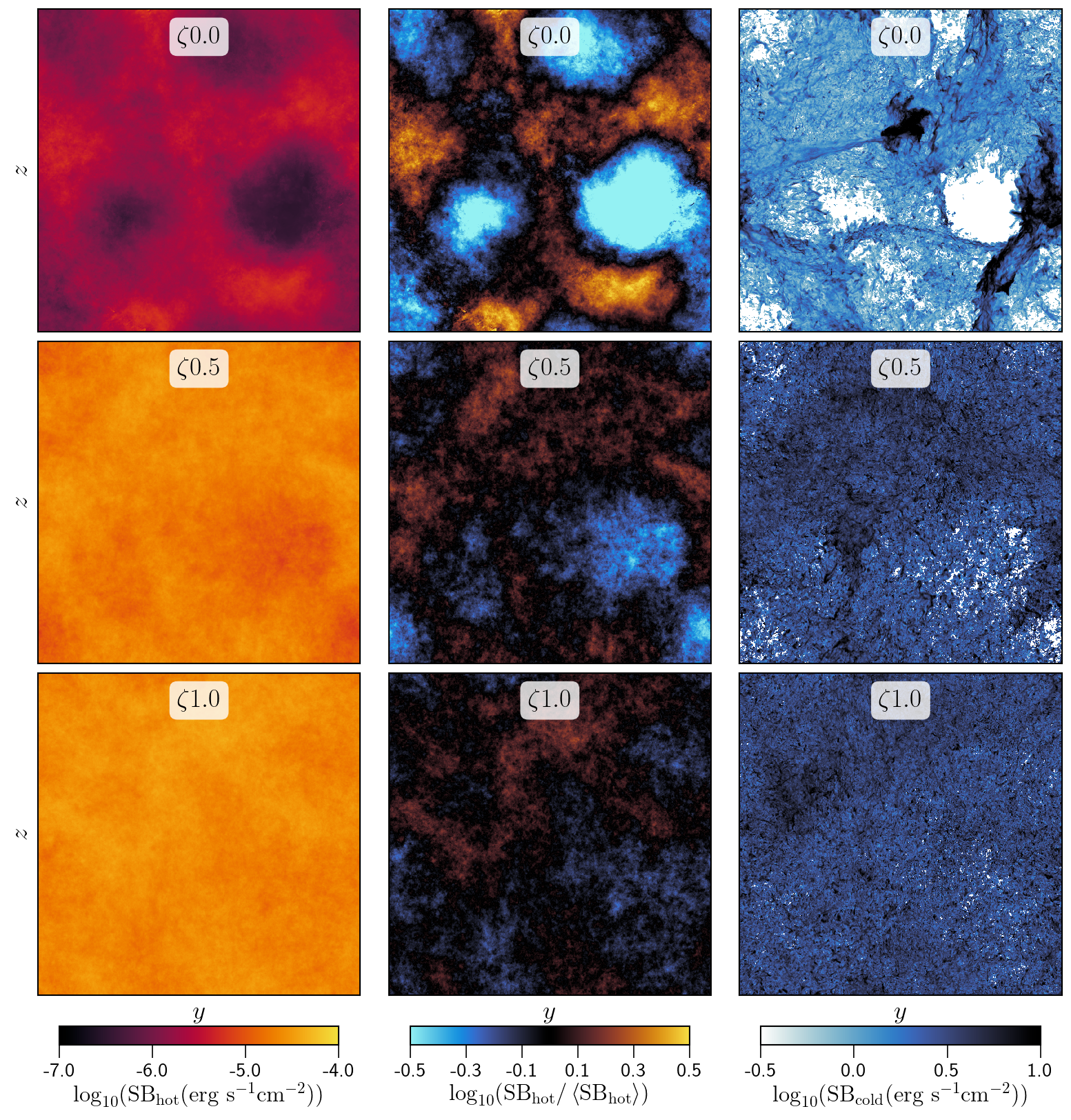}	
	\caption[emission-projection plots]{First column: snapshots of surface brightness along the $x$ direction for gas with $T>10^7$~$\mathrm{K}$ at $t=1.042$~$\mathrm{Gyr}$ for three representative simulations: $\zeta0.0$, $\zeta0.5$ and $\zeta1.0$. Second column: Fluctuations in the hot-phase surface brightness from the mean value.
	Third column: same as first column, but for cold-phase gas with $10^4~\mathrm{K}<T<2\times10^4~\mathrm{K}$. All colourbars are in log scale, and are shown at the bottom of each column. The amplitudes of emission from both hot and cold phases increase with increasing $\zeta$. However, the amplitude of relative surface brightness fluctuations follows the opposite trend, decreasing with increasing $\zeta$, due to the decreasing effect of large-scale compressive modes, which cause the large fluctuations. We find a strong spatial correlation between the large-scale variations of hot- and cold-phase surface brightness variations.}
	\label{fig:emission-proj-2d}
\end{figure*}

In this subsection, we discuss some observational implications of the turbulence driving. In \cref{fig:emission-proj-2d}, we show the mock projected emission (integrated along the $x$-axis) from the X-ray emitting hot phase ($T\gtrsim10^7~\mathrm{K}$; first column, denoted by $\mathrm{SB}_{\mathrm{hot}}$), normalised surface brightness fluctuations in the hot phase (second column) and projected emission from the H$\alpha$-emitting cold phase ($10^4~\mathrm{K}<T<2\times10^4~\mathrm{K}$; third column, denoted by $\mathrm{SB}_{\mathrm{cold}}$) for our fiducial set of runs. 

\subsubsection{Mock hot-phase emission}\label{subsubsec:hot_phase_emission}
In the first column of \cref{fig:emission-proj-2d}, all three runs show diffuse and volume-filling hot-phase emission along all sight lines. The net X-ray surface brightness ($\mathrm{SB}_{\mathrm{hot}}$) is the smallest for the $\zeta0.0$ run and increases with increasing $\zeta$. The emission depends on density and temperature as $\sim\rho^2\sqrt{T}$ for Bremsstrahlung emission which dominates at these temperatures. The temperature of the hot phase decreases with increasing $\zeta$, whereas its density increases, as seen in figures~\ref{fig:temp-PDF} and \ref{fig:dens-PDF}, respectively. The overall trend thus follows the trend in density, due to its stronger weight. The brightness shows large spatial variations for the $\zeta0.0$ run, varying over almost two orders of magnitude, whereas it is more uniform for the larger $\zeta$ runs. We observe large-scale X-ray cavities in the hot-phase emission, which suggests that the driving by expanding AGN jet-inflated bubbles could be compressive in nature.

\subsubsection{Mock surface brightness fluctuations}\label{subsubsec:X-ray_SB_fluc}
The normalised X-ray surface brightness fluctuations ($\log_{10}(\mathrm{SB}_{\mathrm{hot}}/\langle\mathrm{SB}_{\mathrm{hot}}\rangle)$, shown in the second column of \cref{fig:emission-proj-2d}) have been studied in several recent observations, such as
\cite{churazov2012x,zhuravleva2014turbulent,Walker2015MNRAS,zhuravleva2018}. They are directly related to gas-density fluctuations and are used as an indirect method to estimate turbulent velocities of the hot phase\footnote{Note that the observed fluctuations in the X-ray brightness of the ICM are limited by the voxel size, so small scale fluctuations are likely to be smoothed out. The density stratification of the ICM as well as the existence of sub-structures further complicate the calculations of brightness fluctuations, since the mean brightness is a function of the radial and azimuthal co-ordinates. We refer the readers to \cite{Vazza2011MNRAS,Simonte2022A&A} who use filtering methods to remove the effect of substructures while calculating fluctuations in cosmological simulations.}. Although the gas has similar turbulent velocities for all three of our runs (see column~6 of \cref{tab:sim_params}), the amplitude of fluctuations varies significantly across our three fiducial runs. The $\zeta0.0$ run shows strong, large-scale variations in brightness. With increasing $\zeta$, these fluctuations occur at smaller scales and have much smaller amplitudes.
This reflects the trend in gas-density fluctuations that we discussed in \cref{subsubsec:dens_distribution}. On the driving scale, compressive and expansive motions lead to larger density differences compared to solenoidal motions. Further, the high-density regions cool fast, generating contact discontinuities and further increase the amplitude of density and surface brightness fluctuations.

\subsubsection{Mock emission from H$\alpha$ filaments}\label{subsubsec:Mock_Halpha_emission}
We shown the net emission from gas with $10^4~\mathrm{K}<T<2\times10^4~\mathrm{K}$ in the third column of \cref{fig:emission-proj-2d}. These are comparable to the atomic filaments studied in observations such as \cite{hu1992,Conselice2001AJ,Olivares2019A&A,Boselli2019A&A}. We find that in general, the amplitude of $\mathrm{SB}_{\mathrm{cold}}$ increases with increasing $\zeta$, but the $\zeta0.0$ run also shows strong local peaks in $\mathrm{SB}_{\mathrm{cold}}$. These peaks are co-spatial with the peaks in $\mathrm{SB}_{\mathrm{hot}}$ and can be explained by the trends in the density PDF (see \cref{fig:dens-PDF}). The mean density of the cold phase increases with increasing $\zeta$, which explains the general increase in $\mathrm{SB}_{\mathrm{cold}}$ with increasing $\zeta$. At the same time, the $\zeta0.0$ run has a high-density tail, which is reflected in the strong localised emission peaks.
Unlike $\mathrm{SB}_{\mathrm{hot}}$, $\mathrm{SB}_{\mathrm{cold}}$ does not cover all sight lines for any of our simulations. We observe both small-scale and large-scale filaments for the $\zeta0.0$ run--the large-scale features are due to compressive modes and the small-scale features are due to solenoidal modes, which form cold gas through turbulent mixing of the hot and cold phases. The sightline/sky-covering fraction increases with increasing $\zeta$ due to the decreasing effect of the large-scale expansive motions and due to efficient mixing. 

For the $\zeta0.0$ run, we find strong cold-phase emission surrounding the hot-phase cavities (in the first column) and lack of any emission from the regions inside the cavities. This feature can be explained by two mechanisms--(i) the pre-existing cold-phase gas is swept up outwards by the expanding hot cavities, and (ii) the cold gas forms through condensation from the swept-up dense, fast-cooling regions of the hot phase. Both of these mechanisms take place simultaneously in our simulations (see temperature projection movie in supplementary material). These features are commonly reported in observations, where cold (atomic and molecular) gas filaments are found to surround X-ray cavities and/or are coincident with bright X-ray peaks \citep[for example,][]{David2017ApJ,Ricci2018ApJ,OSullivan2021MNRAS}. Similar to our simulations, these filaments could be transient features caused by large-scale compressive motions.


\subsection{Scale-dependent turbulence statistics}\label{subsec:scale_statistics}
In this subsection, we discuss two scale dependent turbulent statistics--(i) the second-order velocity structure function (denoted by $\mathrm{VSF}_2$), and (ii) the spatial cross-correlation function (denoted by $\mathrm{VCF}(r)$). In order to study the scale dependence of compressive and solenoidal modes, we have decomposed $\mathrm{VSF}_2$ into its transverse and longitudinal components $\mathrm{VSF}_{2,\mathrm{trsv}}$ and $\mathrm{VSF}_{2,\mathrm{long}}$, respectively. These quantities are defined as
\begin{subequations}
\begin{align}
    &\mathrm{VSF}_2(r) = \langle \delta v_r^2 \rangle,\label{eq:VSF_2}\\ 
    &\mathrm{VSF}_{2,\mathrm{trsv}}(r) = \langle \delta v_{\mathrm{trsv},r}^2 \rangle,\label{eq:VSF_2_trsv}\\ 
    &\mathrm{VSF}_{2,\mathrm{long}}(r) = \langle \delta v_{\mathrm{long},r}^2 \rangle\text{, where}\label{eq:VSF_2_long}\\ 
    &\delta v_r = |\bm{v}(\bm{x}+\bm{e}_1r,t) - \bm{v}(\bm{x},t)|,\label{eq:delv_VSF}\\
    &\delta v_{\mathrm{long},r}=\left(\bm{v}(\bm{x}+\bm{e}_1r,t) - \bm{v}(\bm{x},t)\right)\cdot\bm{e}_1\text{, and}\label{eq:delv_long_VSF}\\
    &\delta v_{\mathrm{trsv},r}=\left(\delta v_r^2-\delta v_{\mathrm{long},r}^2\right)^{1/2}.\label{eq:delv_trsv_VSF}
\end{align}
The spatial cross-correlation function between hot- and cold-phase gas gives us an estimate of the coupling between the two phases as a function of their separation. It is defined as
\begin{equation}
\mathrm{VCF}_{\mathrm{hc}}(r)=\langle \bm{v}_{\mathrm{cold}}(\bm{x})\cdot\bm{v}_{\mathrm{hot}}(\bm{x}+\bm{e_1}r)\rangle.    \label{eq:vcorrfunc_cold_hot}
\end{equation}
For the cross-correlation function, we choose the two points $\bm{x}$ and $\bm{x}+\bm{e_1}r$ such that $\bm{x}$ lies in the cold phase ($10^4~\mathrm{K}<T<2\times10^4~{\mathrm{K}}$) and $\bm{x}+\bm{e_1}r$ lies in the hot phase ($T>10^7~\mathrm{K}$).
\end{subequations}
We discuss these scale-dependent turbulent statistics for our fiducial runs and their implications for the gas in galactic halos below.

\subsubsection{Second-order velocity structure function}\label{subsubsec:VSF2}
\begin{figure*}
		\centering
	\includegraphics[width=2.0\columnwidth]{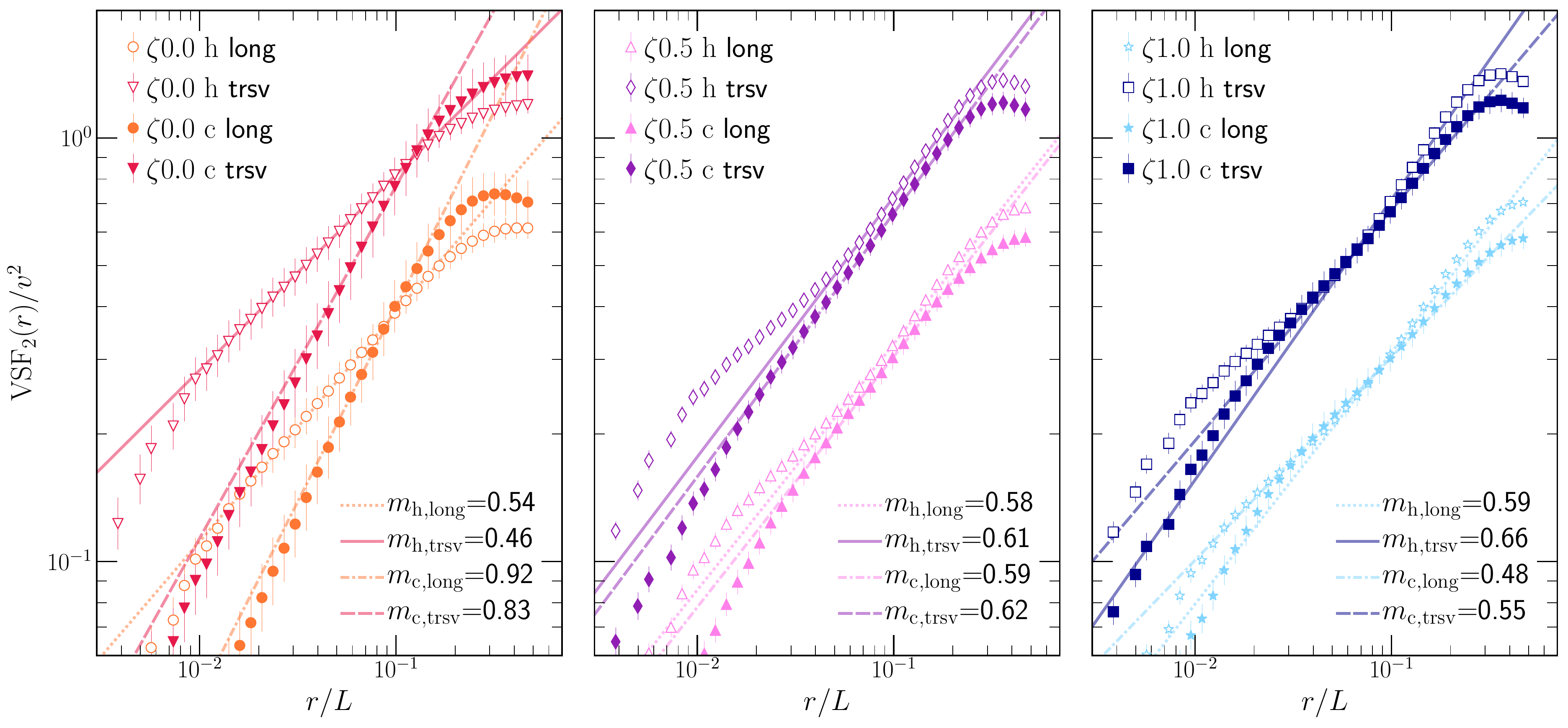}	
	\caption[density PDF]{Second-order structure function of hot ($T > 10^7 K$, unfilled data points) and cold phase $T < 2 \times 10^4 K$, filled data points) gas, showing the transverse and longitudinal components, for our three representative runs. The $\mathrm{VSF}_2$ are normalised by $v^2$. The error bars denote the temporal variation in the $\mathrm{VSF}_2$. We fit separate power-law scaling relations to the $\mathrm{VSF}_2$ of the two phases, as well as to the transverse and longitudinal components. We denote the scaling exponents by $m$ and the phases by the subscripts $h$ and $c$, and the components by $\mathrm{trsv}$ and $\mathrm{comp}$. These slopes are listed in the bottom right part of each panel.
	The transverse and the longitudinal components have similar scaling with separation $r$, but the transverse component has a larger magnitude for all our runs. The cold-phase $\mathrm{VSF}$ is steeper than the hot-phase $\mathrm{VSF}$ for $\zeta=0.0$ but it becomes parallel to the hot phase for $\zeta=0.5$ and shallower for $\zeta=1.0$.}
	\label{fig:vsf2_hc_3d}
\end{figure*}

In \cref{fig:vsf2_hc_3d}, we show the second-order velocity structure function for the hot and cold phases for our fiducial set of runs. The velocity structure functions represent the velocity difference between two points as a function of separation and are useful for determining turbulence statistics such as (i) the driving scale(s) of turbulence--indicated by a flattening of $\mathrm{VSF}$ at the corresponding scale(s); (ii) the distribution of turbulence kinetic energy across scales and comparison with theoretical predictions such as Burgers turbulence \citep{Burgers1948171} and Kolmogorov turbulence \citep{kolmogorov1941dissipation}. Recently, integral field spectroscopy observations of atomic filaments \citep{Gendron-Marsolais2018MNRAS,Sarzi2018MNRAS,Tremblay2018ApJ,Boselli2019A&A} and Atacama Large Millimeter Array (ALMA) observations of molecular gas filaments \citep{Simionescu2018MNRAS,Tremblay2018ApJ} have been used to determine the $\mathrm{VSF}$ for the cold phase of the ICM \citep{Li2020ApJ}. Both observations and recent numerical simulations such as \cite{Hillel2020ApJ,Wang2021MNRAS,Gronke2022MNRAS,Mohapatra2022VSF,Hu2022arXiv,Zhang2022arXiv} have attempted to interpret the slopes and amplitude of the cold phase $\mathrm{VSF}$ and its implications for the turbulence kinematics of the hot phase.
Using our multiphase turbulence simulations, we study the $\mathrm{VSF}_2$ of the hot and cold phases, as well as the distribution of the kinetic energy into compressive and solenoidal modes. 

We normalise the $\mathrm{VSF}_2$ by $v^2$ and decompose it into its transverse and longitudinal components. For comparison with non-radiative runs with the same $\zeta$, we present the decomposed $\mathrm{VSF}_2$ for our three non-radiative runs in \cref{fig:app_vsf2_nocool}. For all runs and both phases, both the longitudinal and transverse components of the $\mathrm{VSF}_2$ increase as a function of $r$ and peak close to the driving scale. The transverse component has a larger amplitude compared to the longitudinal component for all runs and both phases, as expected from the results in \cref{subsec:comp_sol_ratio}. 


For the $\zeta0.0$ run, both the longitudinal and transverse components of the cold-phase $\mathrm{VSF}_2$ are larger than the corresponding hot phase $\mathrm{VSF}_2$ at the driving scale ($L/2$). However, they are much steeper than the hot-phase $\mathrm{VSF}_2$, and on small scales, the hot-phase $\mathrm{VSF}_2$ is much larger. The large cold-phase velocities close to the driving scale could be due to strong compressive motions associated with multiphase condensation of high-density regions (notice that $\mathrm{VSF}_2^{\mathrm{cold}}/\mathrm{VSF}_2^{\mathrm{hot}}$ is much larger for the longitudinal component), in addition to the external turbulence driving. 
The steepness of the structure function could be due to Burgers turbulence ($\mathrm{VSF}_2\propto\ell$, \citealt{Burgers1948171}) in the large, supersonic cold clouds. The $\mathrm{VSF}_2$ could also be affected by weak coupling between the hot and cold phases due to large $\chi$ (see \cref{fig:dens-PDF}), which in turn is caused by the large-scale compressive driving.

With increasing $\zeta$, both components of the hot-phase $\mathrm{VSF_2}$ become steeper, whereas the cold-phase $\mathrm{VSF_2}$ becomes flatter. For the hot phase, this trend is the opposite of what we observe in the non-radiative runs (see appendix A, subsection 2), where the $\mathrm{VSF}_2$ flattens upon increasing $\zeta$, as is expected from turbulence scaling changing from compressible Burgers-like to incompressible Kolmogorov-like with increasing $\zeta$. The flatter $\mathrm{VSF}_2$ is likely due to the contribution of velocities due to thermal instability, which are expected to scale as $r^0$ since both small- and large-scale isobaric modes are expected to grow at the same rate, as we discussed in section~4.3.1.2 in \cite{Mohapatra2022VSF} \citep[also see fig.~2 of ][which shows similar growth rate of small- and large-scale isobaric modes]{sharma2010thermal}. The flattening of velocity scaling due to thermal instability is also observed in the velocity power spectra of \cite{Gazol2010ApJ}, where the velocity power spectrum for their non-isothermal run with thermal instability has a flatter scaling compared to their isothermal run. The velocities due to thermal instability are expected to be larger for the compressive driving run, since the driving generates stronger density fluctuations, which act as seeds for the isobaric modes to grow from. The flattening of the cold-phase $\mathrm{VSF}_2$ with increasing $\zeta$ is expected due to two reasons-- (i) smaller velocities on large scales, due to weaker contribution from velocities associated with large-scale gas condensation and (ii) larger cold-phase velocities at intermediate scales ($0.05\lesssim r/L\lesssim0.1$) due to stronger coupling between the hot and cold phases with decreasing $\chi$.


\subsubsection{Velocity cross-correlation function}\label{subsubsec:VCF}

\begin{figure}
		\centering
	\includegraphics[width=\columnwidth]{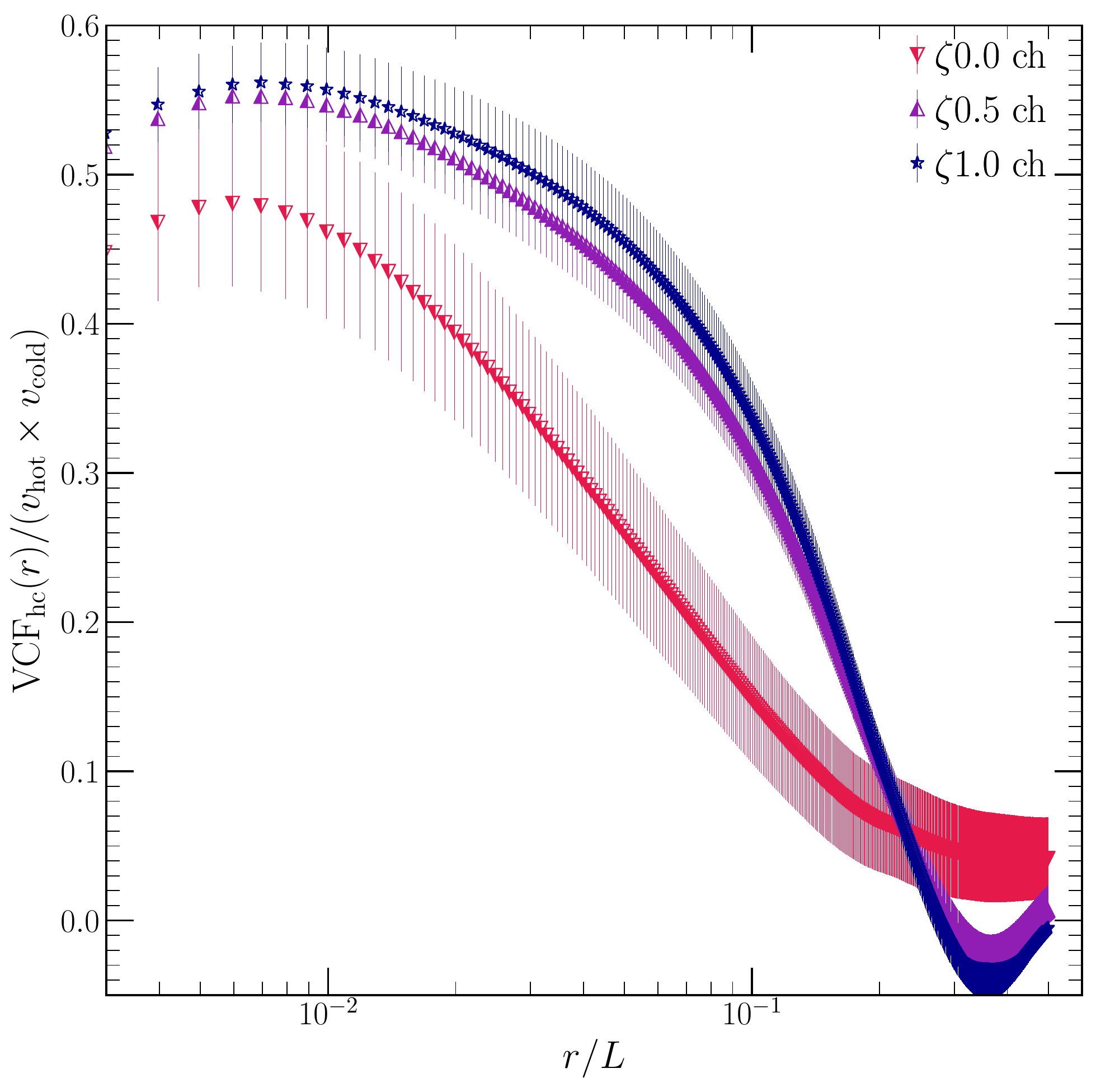}	
	\caption[hot-cold correlation function]{Cross correlation function between hot- and cold-phase velocities, for our three representative runs. The correlation between the velocities decreases with increasing $r$ and increases with increasing $\zeta$.}
	\label{fig:vcorrf2_hc_3d}
\end{figure}

In \cref{fig:vcorrf2_hc_3d}, we show the spatial cross-correlation function between the hot- and cold-phase velocities $\mathrm{VCF}_{\mathrm{hc}}(r)$, as defined in \cref{eq:vcorrfunc_cold_hot}. The $\mathrm{VCF}_{\mathrm{hc}}$ is a useful statistical tool to check the correlation between gas velocities as a function of separation and it gives us a sense of coupling between the hot and cold phases. We have normalised $\mathrm{VCF}_{\mathrm{hc}}(r)$ by dividing out the standard deviations of hot- and cold-phase velocities $v_{\mathrm{hot}}$ and $v_{\mathrm{cold}}$, respectively. For all three fiducial runs, the $\mathrm{VCF}_{\mathrm{hc}}$ peaks on small scales and decreases as a function of separation $r$. With increasing $\zeta$, the value of $\mathrm{VCF}_{\mathrm{hc}}$ increases, due to stronger coupling between the phases. The coupling becomes stronger due to smaller density contrast ($\chi$) in larger $\zeta$ runs, since large-scale compressive motions are weaker. The dependence on $\zeta$ in these runs are in line with our results from \cite{Mohapatra2022VSF}, where we found the amplitude of the $\mathrm{VCF}_{\mathrm{hc}}$ to depend inversely on $\chi$.

\subsection{CGM case}\label{subsec:CGM case}
\begin{figure*}
		\centering
	\includegraphics[width=2.0\columnwidth]{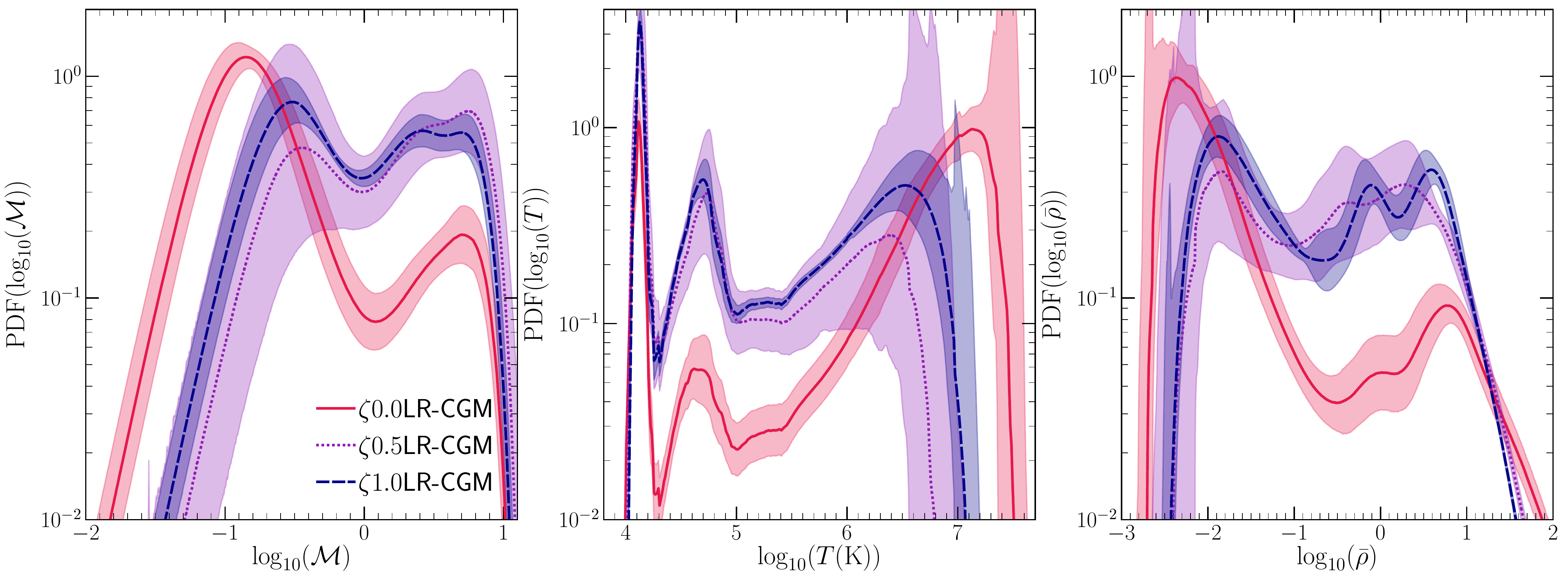}	
	\caption[PDF CGM case]{The volume-weighted PDFs of the logarithms of $\mathcal{M}$ (first column), temperature (second column) and density (third column) for three runs with CGM-like initial conditions. The PDFs are temporally averaged over 26 snapshots from $t=0.98$~$\mathrm{Gyr}$ to $t=1.95$~$\mathrm{Gyr}$, and the shaded region shows the $1\sigma$ variation. With decreasing $\zeta$, the PDFs show more mixing between the hot and cold phases. For the $\zeta0.5$LR-CGM and $\zeta1.0$LR-CGM runs, we observe a strong third peak at intemediate temperatures and densities.}
	\label{fig:CGM_PDFs}
\end{figure*}

In this subsection, we discuss the effects of the turbulence driving on our simulations with CGM-like initial conditions. We focus on its effects on the volume-weighted PDFs of $\mathcal{M}$, temperature (T) and normalised density ($\bar{\rho}$), shown in the first, second and third panels of \cref{fig:CGM_PDFs}, respectively. For all three runs, the hot phase is subsonic, with $10^6~\mathrm{K}\lesssim T\lesssim10^7~\mathrm{K}$ and density $\sim10^{-4}$--$10^{-2}~\mathrm{cm}^{-3}$ ($\bar{\rho}=\rho/\rho_0$, where $\rho_0=0.086~\mathrm{cm^{-3}}$). 

For the $\zeta0.0$-LR-CGM run, the gas distribution is bimodal with strong and distinct peaks corresponding to the two temperature phases, similar to our ICM-like runs in \cref{fig:temp-PDF}. The hot phase is at somewhat higher temperatures due to the strong heating by fast dissipation of compressive modes. The compressive and expansive gas motions also lead to high-density cold-phase gas and lower-density hot-phase gas compared to the higher $\zeta$ runs.

With increasing $\zeta$, the density contrast between hot and cold phases becomes weaker in the absence of large-scale compressive modes. The gas distribution is more continuous rather than bimodal, with the hot and cold peaks moving closer to each other. Since $\chi$ is much smaller in the CGM compared to the ICM case, mixing due to turbulence continuously produces gas at intermediate temperatures  $T_{\mathrm{mix}}\sim\sqrt{T_{\mathrm{hot}}T_{\mathrm{cold}}}\sim10^5~\mathrm{K}$ \citep{Begelman1990MNRAS}. Since $T_{\mathrm{mix}}$ is close to the peak of the \cite{Sutherland1993} cooling curve, this gas has a short cooling time and thus cools rapidly and accumulates just below the peak of the cooling curve, at $\sim10^{4.7}~\mathrm{K}$. This third peak also shows up in the $\mathcal{M}$ and density distributions, as a strong peak at intermediate $\mathcal{M}$ and $\bar{\rho}$. The effect of increasing $\zeta$ on the PDFs is much stronger for the CGM-like runs, since the value of $\chi$ is an order of magnitude smaller than in our ICM-like runs, leading to stronger mixing. \cite{Gronke2022MNRAS} also study the effect of $\chi$ on mixing between hot and cold phases and find turbulent mixing to become stronger with decreasing $\chi$.

In section~3.3.2 of \citetalias{Mohapatra2022characterising}, we highlighted that the amount of gas with temperatures in between the hot and cold phase depends on the turbulent heating fraction $f_{\mathrm{turb}}$. Since we only used solenoidal forcing in the previous study, increasing $f_{\mathrm{turb}}$ led to both a smaller $\chi$ and increased amplitude of solenoidal velocities, both of which are associated with increased mixing between the hot and cold phases and larger amounts of gas at intermediate temperatures. In this study, the effect of increasing $\zeta$ decreases $\chi$ and as a consequence produces larger amounts of gas at intermediate temperatures, even though $v_{\mathrm{sol}}$ is similar across all runs. 

The gas at these temperatures is traced by ions such as Mg\texttt{II} at $\sim10^4~\mathrm{K}$ \citep{Anand2021MNRAS,Anand2022arXiv}, Si\texttt{IV} at$\sim10^{5}~\mathrm{K}$ \citep{Zheng2017ApJ} and  O\texttt{VI} at $\sim10^{5.5}~\mathrm{K}$ \citep{Tumlinson2011ApJ}. We refer the reader to fig.~6 of \cite{Tumlinson2017review} for a larger list of ions that are abundant at these temperatures. These ions are detected as absorption features in the spectra of background quasars as well as emission from nearby bright sources. By comparing the relative abundance of these ions and using our estimate of $\chi$, we can constrain the amplitude of the turbulent velocity. We can use this information further to infer the nature of the driving (compressive vs solenoidal) by the different sources of turbulence, such as stellar winds, supernovae, outflows, mergers and tidal interactions.

\section{Caveats and future work}\label{sec:caveats-future}
In this section, we discuss some of the assumptions and shortcomings of the present study. We also present some future prospects.

We model the ICM as a fluid in a periodic box with a fixed injection scale for turbulence, which allows us to perform a controlled parameter scan and study the dependence of gas properties and kinematics on the turbulence driving. However, the driving due to supernovae, AGN jets, galaxy infall and mergers, etc., can occur across multiple scales and contribute to both solenoidal and compressive modes during the course of a galaxy/cluster's evolution, which are not captured by our simple model. 

In order to run our simulations with the available computing resources, we have used subcycling for the radiative cooling, as explained in paragraph \ref{par:cooling_subcycling}. We set $\mathrm{sub}_{\mathrm{factor}}=25$ for our fiducial set of runs. Implementing subcycling affects the gas density, temperature and pressure distributions, as we discuss in \cref{app:mach_temp_dens_dist_subc}. Larger values of $\mathrm{sub}_{\mathrm{factor}}$ lead to less amount of gas at intermediate temperatures (see \cref{fig:app_subc_dependence}), since this gas has a short cooling time and is cooled all the way to $T_{\mathrm{cutoff}}$ when we evolve the cooling module independent of the hydro evolution. The density cooling cutoff ($\rho_{\mathrm{cutoff}}$, see paragraph~\ref{par:cool_cutoffs}) affects the high-density tail of the density distribution for the compressive driving run. For this reason, we have excluded regions with $\rho>\rho_{\mathrm{cutoff}}$ from our analysis.

We have ignored important physics such as magnetic fields, thermal conduction, gravity and density stratification of the ICM. All of these can affect the physics of gas interactions as well as the evolution of thermal instabilities in the ICM. 

Turbulence kinetic energy can be converted into magnetic energy \citep{Federrath2016JPlPh,DiGennaro2021NatAs}, increase the coupling between hot and cold phases and also affect the slopes of the $\mathrm{VSF}_2$ \citep{Mohapatra2022VSF}. Magnetic fields can also affect the growth rate of enstrophy, as seen in studies by \citet{Porter2015ApJ} and \citet{Seta2022arXiv}. On sub-$\mathrm{kpc}$ scales, the ICM is expected to be weakly collisional. Kinetic plasma instabilities such as the firehose and mirror instabilities can affect the evolution of magnetic fields and other gas properties \citep[see e.g.,][]{Santos-Lima2014ApJ,Santos-Lima2016MNRAS}. 

Thermal conduction has been shown to affect the structure functions/power spectra on small scales \citep{gaspari2013constraining,gaspari2014} and can smear out some of the strong temperature gradients that we observe in the $\zeta0.0$ run. 

Gravity and density stratification can affect the growth of thermal instability depending on the ratio between $t_{\mathrm{cool}}$ and the free fall time scale \citep{sharma2012thermal,choudhury2016,Voit2021ApJ}. It can provide another channel of energy transport, from turbulence kinetic energy into  buoyancy potential energy \citep{Mohapatra2020}. In the presence of gravity and the absence of pressure support, the cold-phase filaments may fall towards the cluster centre, as seen in \cite{Wang2021MNRAS}. We plan to include the effects of gravity and magnetic fields in a future study.

We have modelled external, non-turbulent sources of energy by distributing the heat throughout the volume as a function of the local gas density. In future works, we plan to implement heating due to other feedback sources such as cosmic rays, which are expected to have a different distribution of energy and momentum compared to our simple model \citep{Ji2020MNRAS,Butsky2020ApJ,Butsky2021arXiv}.

Our simulations have a resolution of $\sim40~\mathrm{pc}$ for the ICM-like runs and $\sim10~\mathrm{pc}$ for the CGM-like runs. The minimum cooling length $\ell_{\mathrm{cool}}=\min(c_st_{\mathrm{cool}})\approx0.1~\mathrm{pc}$, which is one of the 
smallest length scale in the problem and perhaps necessary for resolving the turbulent mixing layers between the hot and cold phases (\citealt{mccourt2018}; but bottom right panel in Fig. 17 of \citealt{Das2021MNRAS} shows that clouds even smaller than this length scale condense in 1D). This scale is still two orders of magnitude smaller than our current resolution. We refer the reader to \cref{app:mach_temp_dens_dist_res} for a convergence study. With increasing resolution, we expect the cold gas to form more small-scale misty clouds, especially for the higher $\zeta$ runs. We also expect $\chi$ to decrease and turbulent mixing to be stronger as a consequence. 

\section{Summary and conclusions}\label{sec:Conclusion}
In this work, we have studied the effects of the turbulence driving--compressive ($\zeta0.0$, curl-free), natural mixture ($\zeta0.5$) and solenoidal ($\zeta1.0$, divergence-free) on the different statistical properties of multiphase turbulence and discussed their implications for the ICM and the CGM. Here we present some of the main takeaway points of this work:

\begin{itemize}
    \item Turbulence driving directly affects gas properties on large scales. We observe larger contrasts between the density and temperature of the hot and cold phases with compressive driving compared to solenoidal driving. 
    
    \item The density contrast ($\chi$) between the cold and hot phases plays a key role in several other statistical properties of the gas. A smaller ratio of the cold-to-hot-phase gas density, $\chi=\rho_{\mathrm{cold}}/\rho_{\mathrm{hot}}$, in solenoidal driving leads to more efficient turbulent mixing between the two phases, resulting in larger amounts of gas at intermediate temperatures. This effect is stronger for the CGM-like runs, where $\chi$ is an order of magnitude smaller than in the ICM-like runs.
    
    \item Gas density and pressure fluctuations, which are important observational tools to estimate turbulent gas velocities of the ICM depend on $\zeta$. The density fluctuations in the hot X-ray-emitting phase are orders of magnitude larger than known scaling relations due to additional compressive velocities and contact discontinuities induced by thermal instability of the medium.  Gas-pressure fluctuations scale as $\mathcal{M}^2$, and their amplitudes decrease with increasing $\zeta$. 
    
    \item The gas velocities are dominated by solenoidal modes at all temperatures, even for compressive driving. Baroclinicity efficiently generates enstrophy (vorticity squared), especially for purely compressive driving, which is further amplified by vortex stretching. In steady state, the enstrophy is nearly independent of the driving mode.
    
    \item Compressive driving affects the appearance of the gas in X-rays and optical mock-emission maps, with the X-ray emission showing cavities surrounded by strong emission regions. The H$\alpha$ filaments coincide with the peaks in X-ray emission. The emission maps are relatively uniform for solenoidal driving and show no strong large-scale features.
    
    \item The coupling between the velocities of the hot and cold phases increases with decreasing density contrast ($\chi$), which decreases with increasing fraction of solenoidal driving. For the compressive driving run, the cold-phase 2nd-order velocity structure function, $\mathrm{VSF}_2$, has larger velocities, but has a much steeper scaling with separation (Burgers-like) compared to the hot phase. For all runs, the transverse component of $\mathrm{VSF}_2$ is much larger than the longitudinal component, implying the dominance of solenoidal velocities at all scales for both hot and cold phases.
\end{itemize}

\section*{Acknowledgements}
This work was carried out during the ongoing COVID-19 pandemic. The authors would like to acknowledge the health workers all over the world for their role in fighting in the frontline of this crisis. 
The authors would like to thank the anonymous referee for a constructive report, which helped to improve this work.
CF acknowledges funding provided by the Australian Research Council (Future Fellowship FT180100495), and the Australia-Germany Joint Research Cooperation Scheme (UA-DAAD).
PS acknowledges a Swarnajayanti Fellowship (DST/SJF/PSA-03/2016-17) and a National Supercomputing Mission (NSM) grant from the Department of Science and Technology, India.
We further acknowledge high-performance computing resources provided by the Leibniz Rechenzentrum and the Gauss Centre for Supercomputing (grants~pr32lo, pr48pi and GCS Large-scale project~10391), the Australian National Computational Infrastructure (grant~ek9) in the framework of the National Computational Merit Allocation Scheme and the ANU Merit Allocation Scheme. The simulation software, \texttt{FLASH}, was in part developed by the Flash Centre for Computational Science at the Department of Physics and Astronomy of the University of Rochester.

This work used the following software/packages:
\texttt{FLASH} \citep{Fryxell2000,Dubey2008}, \texttt{matplotlib} \citep{Hunter4160265}, \texttt{cmasher} \citep{Ellert2020JOSS}, \texttt{scipy} \citep{Virtanen2020}, \texttt{NumPy} \citep{Harris2020}, \texttt{h5py} \citep{collette_python_hdf5_2014}, \texttt{LMfit} \citep{Newville2016ascl} and \texttt{astropy} \citep{astropy2018}.

\section{Data Availability}
All relevant data associated with this article is available upon reasonable request to the corresponding author.

\section{Additional Links}
Movies of projected density and temperature of different simulations are available as online supplementary material, as well as at the following links:
\begin{enumerate}
    \item \href{https://youtu.be/qBsJti2R0HY}{Movie} of mass-weighted projection of temperature for the fiducial runs.
    \item \href{https://youtu.be/JvTMXYU4m_U}{Movie} of mass-weighted projection of temperature for the three CGM runs.
    \item \href{https://youtu.be/YXWYFtM0P94}{Movie} of the $\zeta0.0$ simulation.
    \item \href{https://youtu.be/RBjN6RD0k78}{Movie} of the $\zeta0.5$ simulation.
    \item \href{https://youtu.be/aGn7_EX8JvI}{Movie} of the $\zeta1.0$ simulation.
\end{enumerate}



\bibliographystyle{mnras}
\bibliography{refs.bib} 



\appendix
\renewcommand\thefigure{\thesection A\arabic{figure}} 
\setcounter{figure}{0}   
\setcounter{table}{0}   
\section*{Appendix A: Results from non-radiative runs}\label{app:without cooling runs}
In this section, we present some of the results from our non-radiative runs, where cooling is switched off. We have run three non-radiative simulations with $1536^3$ resolution elements, keeping all other initial conditions same as the runs with radiative cooling. These runs are useful to separate between the effects of changing $\zeta$ and switching on radiative cooling.

\subsection{Evolution of enstrophy and its sources in non-radiative runs}\label{subsec:enstrophy_evol_no_cool}
\begin{figure}
		\centering
	\includegraphics[width=\columnwidth]{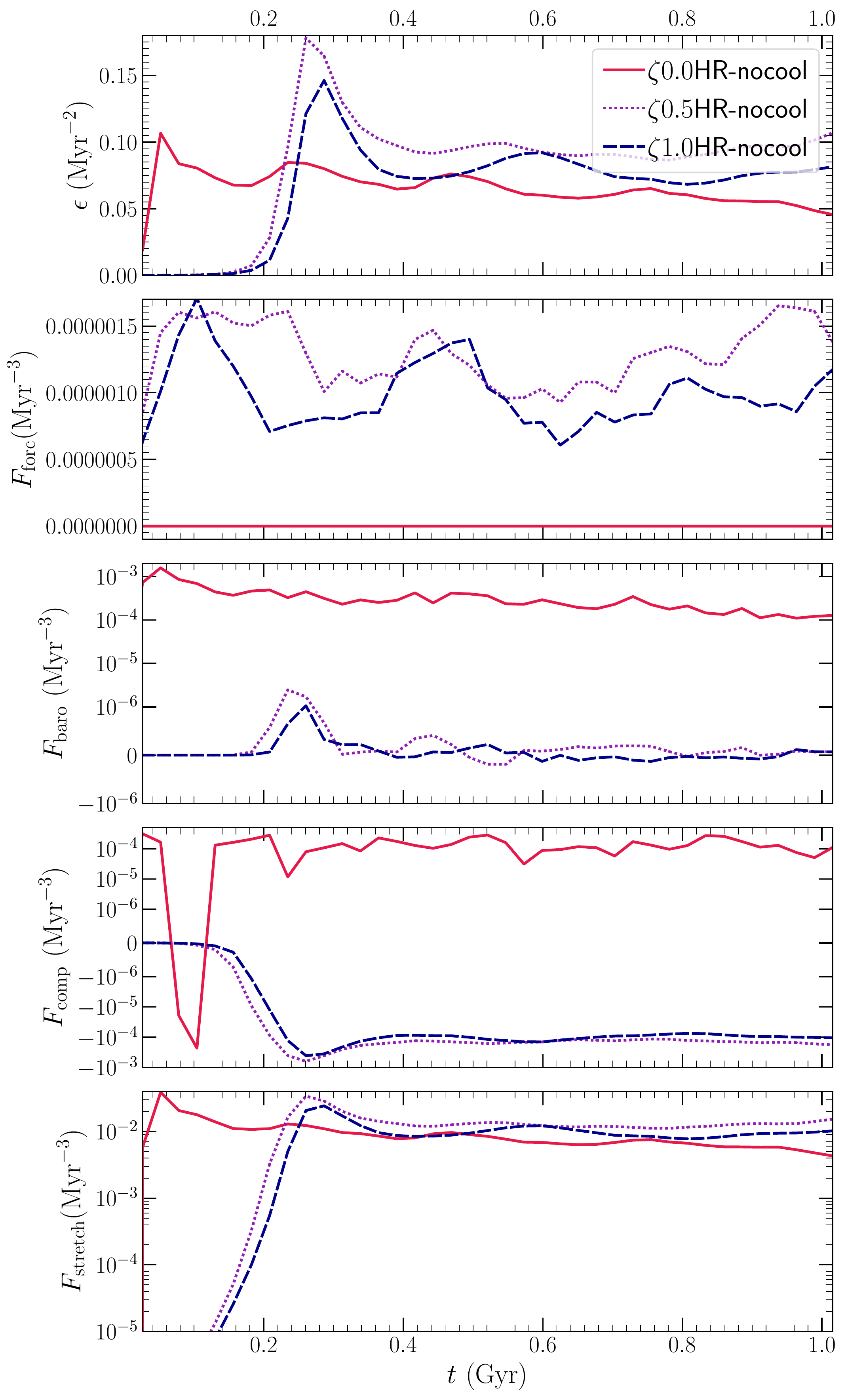}	
	\caption[PDF convergence]{The evolution of enstrophy ($\epsilon$) and its sources for without cooling runs. In the third and fourth panels, the $y$-axis is set to be linear between $-10^{-6}~\mathrm{Myr}^{-3}$ and $10^{-6}~\mathrm{Myr}^{-3}$, and logarithmic otherwise. We discuss the dependence of $\epsilon$ and its sources on $\zeta$ in \cref{app:without cooling runs}.}
	\label{fig:app_enstrophy_nocool}
\end{figure}
In \cref{fig:app_enstrophy_nocool}, we show the evolution of $\epsilon$ and its sources for the three non-radiative runs, similar to \cref{fig:enstrophy-sources-evolution}. Starting from $t=0$, $\epsilon$ shows a sharp increase and reaches a steady state value of the order $0.10~\mathrm{Myr}^{-2}$. Even though the value of $v$ (see col.~6 of \cref{tab:sim_params}) is similar across radiative and non-radiative runs, the amplitude of enstrophy is an order of magnitude smaller compared to the radiative runs. This implies that small scale motions (which dominate the contribution to $\epsilon$) are weaker in the non-radiative runs compared to the radiative runs. 
The $\zeta0.0$HR-nocool run reaches the steady state value within a sound-crossing time, where as the higher $\zeta$ runs take a few $t_{\mathrm{eddy}}$. There is no late evolution stage for the $\zeta0.0$HR-nocool run, unlike the radiative run.
The value of $\epsilon$ has a similar steady state value across all runs, it is slightly larger for the $\zeta0.5$HR-nocool and $\zeta1.0$HR-nocool runs compared to the $\zeta0.0$HR-nocool runs. 

Among the different sources of enstrophy, $F_{\mathrm{forc}}$ is $0$ for the $\zeta0.0$HR-nocool run and subdominant for the other two. 
The baroclinic term is positive and important only for the $\zeta0.0$HR-nocool run. It generates the seeds of enstrophy at $t=0$. In the absence of radiative cooling, and weaker compressive forcing, $F_{\mathrm{baro}}$ is orders of magnitude weaker for the higher $\zeta$ runs, which is expected for weak subsonic solenoidal gas motions. The behaviour of the compression term $F_{\mathrm{comp}}$ is opposite to that of the radiative runs, where the contribution of compressive motions dominates for the $\zeta0.0$HR-nocool run, whereas expansive motions dominate for the larger $\zeta$ runs. This is similar to its role in the radiative runs and it almost cancels out the contribution of $F_{\mathrm{baro}}$ for the $\zeta0.0$HR-nocool run. Similar to the radiative runs, the vortex stretching term is a significant term for enstrophy generation in steady state and is at least an order of magnitude larger than all other source terms. In contrast, the baroclinic term dominated in runs with cooling and heating.

\subsection{Velocity structure functions in non-radiative runs}\label{subsec:VSF_no_cool}
\begin{figure*}
		\centering
	\includegraphics[width=1.8\columnwidth]{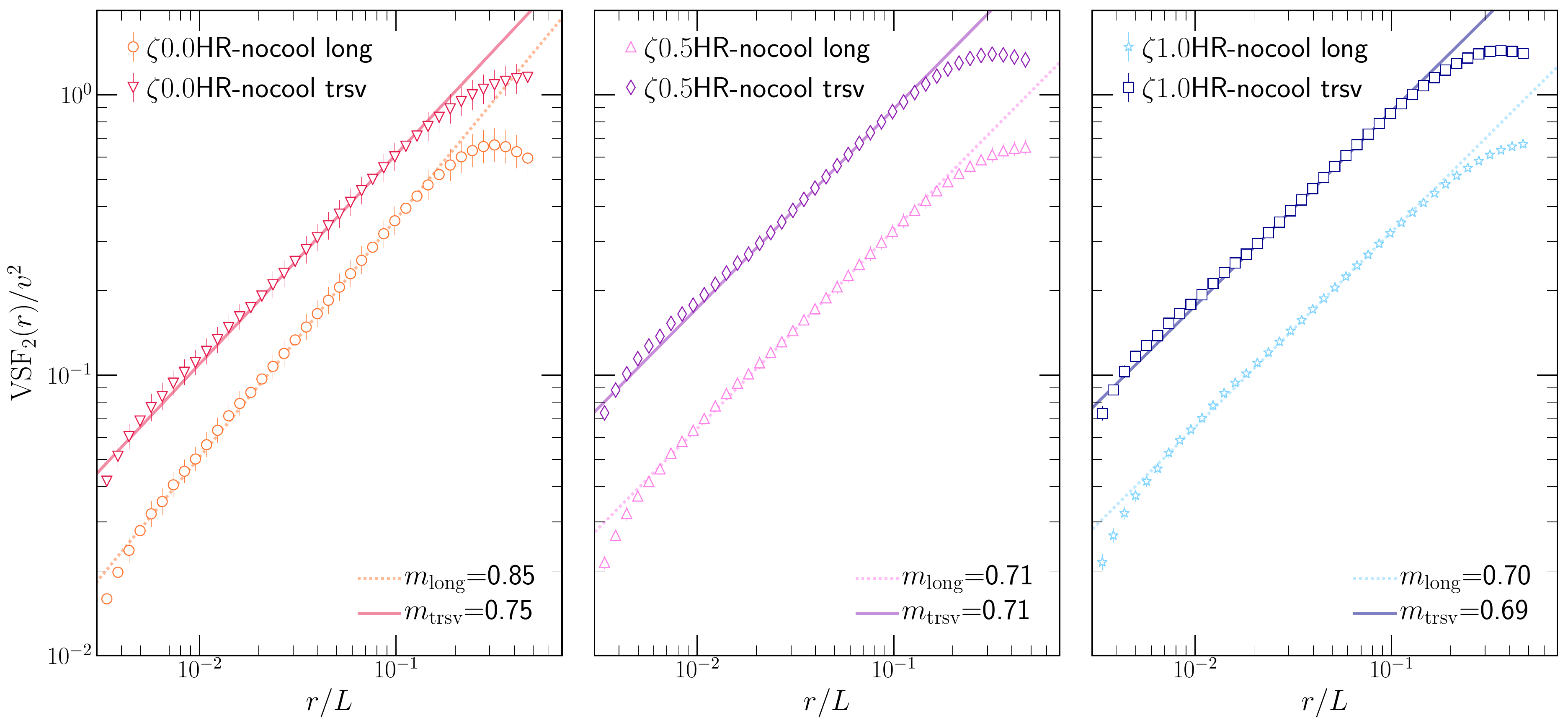}	
	\caption[vsf2 without cooling]{The longitudinal and transverse second order velocity structure functions for our high resolution three non-radiative simulations. The longitudinal component has a smaller amplitude than the transverse component and has a slightly steeper slope. The slopes of both components are steeper for the compressive forcing run compared to the other two.}
	\label{fig:app_vsf2_nocool}
\end{figure*}
In this subsection, we discuss the scaling of the $\mathrm{VSF}_2$ for our non-radiative runs, shown in \cref{fig:app_vsf2_nocool}. For all three runs, the transverse component of the $\mathrm{VSF}_2$ is much larger than the longitudinal component, although this difference is smaller for the $\zeta0.0$ run. Both components of the $\mathrm{VSF}_2$ become shallower upon increasing $\zeta$, denoting a change from Burgers-like compressible turbulence regime to Kolmogorov-like incompressible turbulence regime. For the $\zeta0.0$HR-nocool run, the longitudinal component is closer to the Burgers slope of $r^{1.0}$ compared to the Kolmogorov slope of $r^{2/3}$. The slopes of the $\mathrm{VSF}_2$ for the $\zeta1.0$ run are consistent with the Kolmogorov slopes, with slight steepening due to intermittency \citep{She1994}.

\renewcommand\thefigure{\thesection B\arabic{figure}} 
\setcounter{figure}{0}   
\setcounter{table}{0}   
\section*{Appendix B: Convergence with resolution: distribution functions}\label{app:mach_temp_dens_dist_res}
\begin{figure*}
		\centering
	\includegraphics[width=1.8\columnwidth]{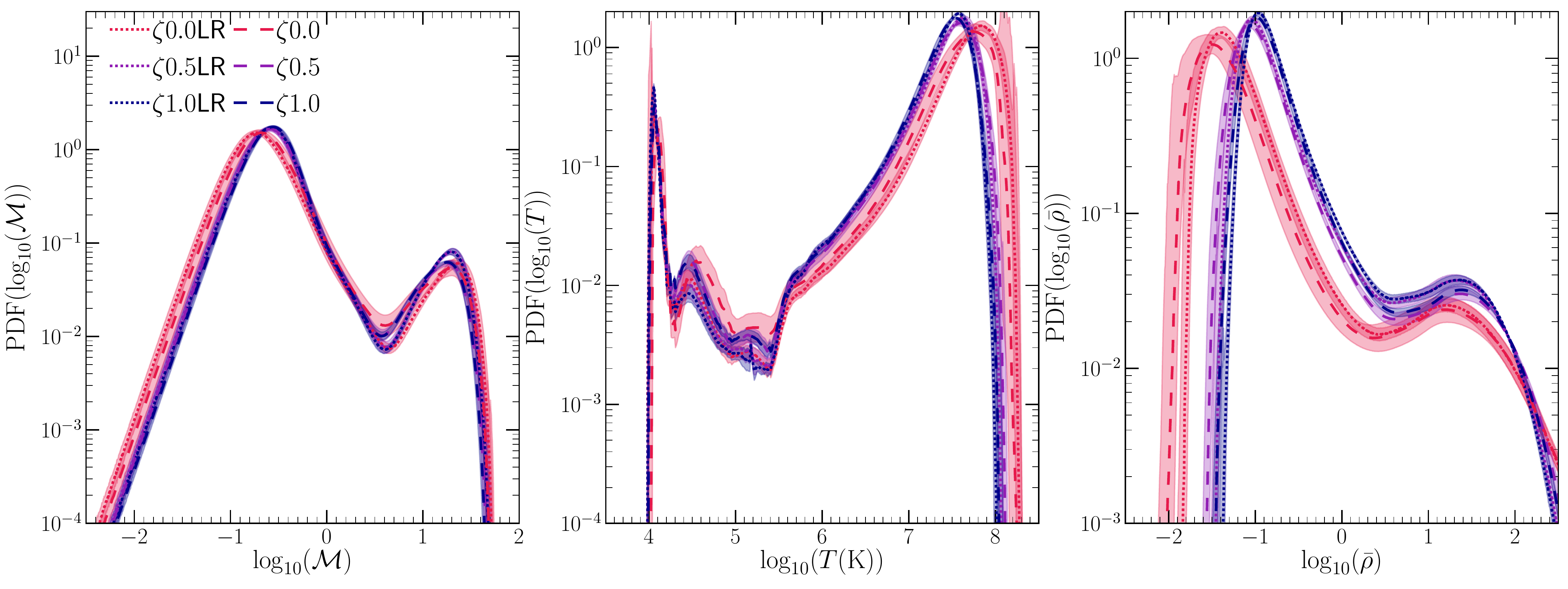}	
	\caption[PDF convergence]{The volume-weighted PDFs of the logarithms of $\mathcal{M}$ (first column), temperature (second column) and density (third column) for our fiducial runs and their lower resolution ($504^3$, labelled LR) counterparts. The PDFs are temporally averaged over 26 snapshots from $t=0.65$~$\mathrm{Gyr}$ to $t=1.3$~$\mathrm{Gyr}$, and the shaded region shows the $1\sigma$ variation. The PDFs do not show a strong dependence on resolution.}
	\label{fig:app_res_conv}
\end{figure*}
In this section, we discuss the convergence of the probability distribution function of different quantities with resolution. In \cref{fig:app_res_conv}. we show the PDFs of the logarithms of $\mathcal{M}$, $T$ and $\bar{\rho}$, respectively. The PDFs are mostly converged with resolution, although we do observe a slight increase in gas at intermediate temperatures ($10^5~\mathrm{K}\lesssim T \lesssim 10^6~\mathrm{K}$) and this is also reflected in increased values of the PDF between the two peaks of the density and Mach number PDFs. This is expected, since with increasing resolution, turbulent mixing due to smaller sized eddies becomes more efficient. Enhanced cooling can contribute to the increased amount of gas at intermediate temperatures, since cold gas can condense down to smaller scales with increasing resolution and not accumulate at a larger scale. For a more detailed discussion on the effect of increasing resolution on these PDFs, we refer the reader to section 3.6.3 of \citetalias{Mohapatra2022characterising}.

\section*{Appendix C: Dependence on $\mathrm{sub}_{\mathrm{factor}}$: distribution functions}\label{app:mach_temp_dens_dist_subc}
\renewcommand\thefigure{\thesection C\arabic{figure}} 
\setcounter{figure}{0}   
\setcounter{table}{0}   
\begin{figure*}
		\centering
	\includegraphics[width=1.8\columnwidth]{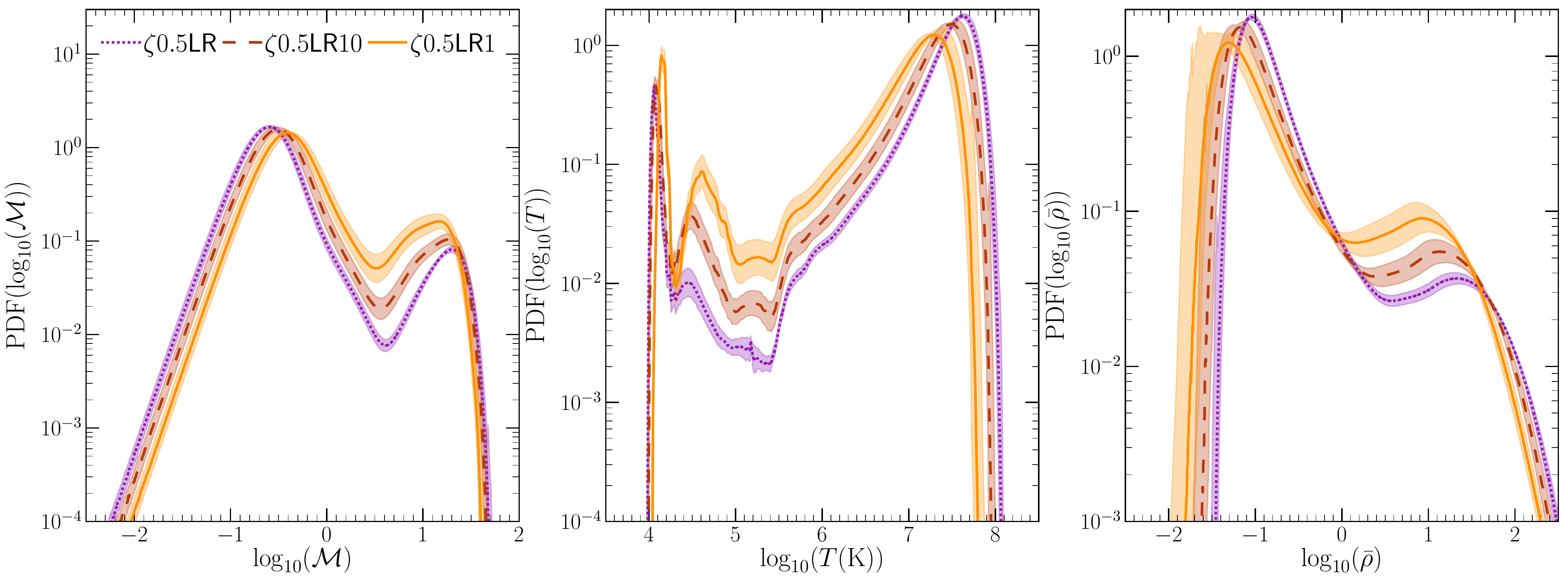}	
	\caption[PDF convergence]{The volume-weighted PDFs of the logarithms of $\mathcal{M}$ (first column), temperature (second column) and density (third column) for three test runs with different values of the subcycling factor $\mathrm{sub}_{\mathrm{factor}}$. The PDFs are temporally averaged over 26 snapshots from $t=0.65$~$\mathrm{Gyr}$ to $t=1.3$~$\mathrm{Gyr}$, and the shaded region shows the $1\sigma$ variation. With decreasing $\mathrm{sub}_{\mathrm{factor}}$, the PDFs show more mixing between the hot and cold phases.}
	\label{fig:app_subc_dependence}
\end{figure*}
In this section, we look for the effect of the cooling subcycling factor $\mathrm{sub}_{\mathrm{factor}}$ (defined in paragraph~\ref{par:cooling_subcycling}) on the Mach number, temperature and density PDFs, shown in \cref{fig:app_subc_dependence}. The default value of  $\mathrm{sub}_{\mathrm{factor}}$ is $25$ for all our simulations. Here, we compare between three runs at $504^3$ resolution with $\mathrm{sub}_{\mathrm{factor}}=25$, $10$ and $1$ (subcycling switched off), respectively. We observe a clear and strong dependence of the PDFs on the value of $\mathrm{sub}_{\mathrm{factor}}$. With decreasing $\mathrm{sub}_{\mathrm{factor}}$, we observe more gas at intermediate temperatures ($10^5~\mathrm{K}\lesssim T \lesssim 10^6~\mathrm{K}$) in all three PDFs. The peaks corresponding to the hot and cold phases in the density and Mach number PDFs also move closer to each other--denoting a decrease in the gas bimodality. This happens because with decreasing $\mathrm{sub}_{\mathrm{factor}}$, the gas cooling per each hydro time-step is slower. Since the cooling time is the shortest for the intermediate temperature gas, for larger values of $\mathrm{sub}_{\mathrm{factor}}$ this gas cools down all the way down to $T_{\mathrm{cutoff}}=10^4~\mathrm{K}$. But when we decrease $\mathrm{sub}_{\mathrm{factor}}$, some gas remains at these intermediate temperatures. This also leads to a slightly lower $\chi$ and enhanced turbulent mixing between the hot and cold phases, which explains all the trends in the PDFs. 

Even though the $\mathrm{sub}_{\mathrm{factor}}$ has a significant effect on $\chi$ and the amount of gas between the hot and cold temperature peaks, the smaller $\mathrm{sub}_{\mathrm{factor}}$ are computationally expensive, since the cooling time is much shorter than the CFL time step making smaller $\mathrm{sub}_{\mathrm{factor}}$ runs is untenable with currently available resources. We aim to address this caveat in future simulations with increased resources.

\bsp	
\label{lastpage}
\end{document}